\documentclass[aps,pra,groupedaddress,amsfonts,amssymb]{revtex4}
\begin{document}
\bibliographystyle{revtex}

\draft                

\def\Tr{{\mbox{tr}}\,}
\def\eqref#1{{(\ref{#1})}}

\title[State estimation for large $N$]{State estimation for large ensembles}

\author{Richard D.~Gill}
\email{gill@math.uu.nl}
\homepage{http://www.math.uu.nl/people/gill}
\altaffiliation{also affiliated to EURANDOM, Netherlands}
\affiliation{Dept.~of Mathematics, University of Utrecht, 
  Box 80010, 3508 TA Utrecht, Netherlands}

\author{Serge Massar}
\email{smassar@ulb.ac.be}
\affiliation{Service de Physique Th\'eorique, CP 225, 
 Universit\'e Libre de Bruxelles, B1050 Brussels, Belgium}

\date{\today}

\begin{abstract}
We consider the problem of estimating the state of a large but finite 
number $N$ of identical quantum systems. As $N$ becomes large the 
problem simplifies dramatically. The only relevant measure
of the quality of estimation becomes the mean quadratic error matrix. 
Here we present a bound on this quantity:
a new quantum Cram\'er-Rao inequality.
The new bound expresses succinctly how in 
the quantum case one can trade information about one parameter for 
information about another. The bound holds for arbitrary
measurements on pure states, but only for separable measurements on
mixed states---a striking example of non-locality without entanglement
for mixed but not for pure states.
Cram\'er-Rao bounds are generally only derived for 
unbiased estimators. Here we give a version of our bound for biased estimators,
and a simple asymptotic version for large $N$.
Finally we prove that when the unknown state belongs to a two 
dimensional Hilbert space our quantum Cram\'er-Rao bound can always be 
attained and we provide an explicit measurement strategy that attains it. 
Thus we have a complete solution to the problem of estimating as 
efficiently as possible the unknown state of a large ensemble of 
qubits in the same pure state. The same is true for qubits in the same
mixed state if one restricts oneself to separable measurements, but
non-separable measurements allow dramatic increase of efficiency.
Exactly how much increase is possible is a major open problem.
\end{abstract}


\maketitle

\section{introduction}\label{introduction}

One of the central problems of quantum measurement theory is the
estimation of an unknown quantum state.
Originally only of theoretical interest, this problem is becoming of
increasing practical importance. Indeed there are now several
beautiful experimental realizations of quantum state reconstruction
in such diverse systems as quantum optics\cite{smithey+93},
molecular states\cite{dunn+95}, trapped ions\cite{leibfried+96} and 
atoms in motion\cite{kurtsiefer+97}.

The theoretical work which is the basis for these experiments
is concerned with devising measurement strategies that 
are simple to realize experimentally and which allow an unambiguous
reconstruction of the quantum state. The best known such technique is
quantum state tomography\cite{risken+89}, adapted in \cite{Leonhardt} 
for the case of finite dimensional
Hilbert spaces. However other techniques are also available, see 
\cite{weigert} for a recent discussion in the case of finite dimensional
Hilbert spaces. However, all these works suppose that
the measurements are perfect and that any operator can be measured
with infinite precision. But in general the quality of the
reconstruction will be limited by experimental error\cite{hradil} or
by finite statistics. The present work is devoted to studying this latter
aspect when the unknown state belongs to a finite
dimensional Hilbert space.

Thus the setting of the problem is that we may dispose of a finite number
$N$ of copies of an unknown quantum state $\rho$ (pure or mixed).
Our task is to determine $\rho$ as well as possible. This is by now a classical 
problem\cite{Helstrom}\cite{Holevo}. 

A common approach is first to specify a cost function which numerically
quantifies the deviation of the estimate from the true state.
One then tries to devise a measurement and estimation strategy
which minimizes the mean cost. Since the mean cost typically depends
on the unknown state itself, one typically averages over all possible 
states to arrive at a single number expressing the quality of the estimation.
However optimal strategies have only been found in some simple highly
symmetric cases (the covariant measurements of \cite{Holevo}, see also 
\cite{MassarPopescu}\cite{mixedS}).

However when the number of copies $N$ becomes large one can hope that
the problem becomes simpler so that one might be able to 
find the optimal strategies in this limit.
The reason for this is that in 
the large $N$ limit the estimation problem ceases to be a `global' problem and
becomes `local'. Indeed for small $N$ the estimated state will
often be very different from the true state. Hence the optimal
measurement strategy must take into account the behaviour of
the cost function for large estimation errors. On the other hand in
the limit of an infinite number of copies any two states can be distinguished
with certainty. So the relevant question to ask about the estimation
strategy is at what {\em rate\/} it distinguishes neighboring states.
And in that case we are only concerned with the behaviour of the estimator 
and of the cost function very close to the true value.

To formulate the problem with precision, let us suppose that the
unknown state $\rho(\theta)$ depends on a vector of $p$ unknown real parameters
$\theta=(\theta_1,\dots, \theta_p)$. For instance the $\theta_i$ could
correspond to various settings or physical properties of the apparatus that produces
the state $\rho$.
After carrying out a measurement on the $N$ copies of $\rho$,
one will guess what is $\theta$. Call
$\hat \theta^N=(\hat\theta^N_1,\dots,\hat\theta^N_p)$ the guessed value.
For a good estimation strategy we 
expect the mean quadratic error (m.q.e.)\ to decrease as $1/N$:

\begin{equation}
\mbox{E}_\theta \left(
(\hat \theta^N_i - \theta_i) (\hat \theta_j^N - \theta_j) \right)
\equiv V_{ij}^N(\theta)\simeq { W_{ij} (\theta) \over N}
\label{V}
\end{equation}
where the scaled m.q.e.\ matrix
$W(\theta)=(W_{ij}(\theta))\simeq N V^N(\theta)$ does not depend on $N$. 
$\mbox{E}_\theta$ denotes the mean taken over repetitions of the
measurement with the value of $\theta$ fixed.

Consider now a smooth cost function $f(\hat \theta, \theta)$, which
measures how much the estimated value $\hat \theta$ 
differs from the true value $\theta$ of the parameter. $f$
will have a minimum at $\hat \theta = \theta$, hence can be expanded as

\begin{equation}
f(\hat \theta, \theta)= f_0(\theta)+\sum_{ij} C_{ij}(\theta) 
(\hat \theta_i - \theta_i) (\hat \theta_j - \theta_j)
+ O(\|\hat \theta - \theta\|^3)
\end{equation}
where $C(\theta)=(C_{ij}(\theta))$ is a nonnegative matrix.
Thus for a reasonable estimation strategy the 
mean value of the cost will decrease as 

\begin{equation}
\mbox{E}_\theta({f}(\hat\theta^N,\theta)) = f_0(\theta) 
+ N^{-1}\sum_{ij} C_{ij}(\theta)W_{ij}(\theta)
+ o(N^{-1})
\label{meancost}
\end{equation}
since we expect the expectation value of higher order terms in 
$\hat\theta - \theta$ to decrease faster then $1/N$.
The  problem has become local: 
only the quadratic cost matrix $C(\theta)$ and the scaled 
mean quadratic error matrix $W(\theta)$ at $\theta$ intervene. 
The essential question about 
state estimation for large ensembles is therefore
{\it what scaled m.q.e.\  matrices 
$W(\theta)$ are attainable through arbitrary measurement and estimation 
procedures\/}?
In  particular, what does the boundary of this set of attainable 
m.q.e.\ matrices look like?

In the case when the parameter $\theta$ is one-dimensional ($p=1$)  the
problem has been solved:
a bound on the variance of unbiased estimators---the quantum 
Cram\'er-Rao bound---was given in \cite{Helstrom}, and a 
strategy for attaining the 
bound in the large $N$ limit was proposed in \cite{BarndorffNielsenGill}.  
This justifies taking the bound to induce
a `distinguishability metric' on the space of
states \cite{Wootters}\cite{BraunsteinCaves}. 
In the case of a multidimensional parameter however, though different 
bounds for
the matrix $W$ have been established, in general they are not 
tight \cite{Helstrom}\cite{YuenLax}\cite{Holevo}.

In this paper we present
a new bound for $W$ in the multiparameter case which is inspired by the
discussion in \cite{BarndorffNielsenGill}.
This bound expresses in a natural way how one can trade information about
one parameter for information about another. The interest of this new
bound depends on the precise problem one is considering:

\begin{itemize}

\item 
When $\rho(\theta)=|\psi(\theta)\rangle\langle\psi(\theta)|$ 
is a pure state belonging to a 2 dimensional
Hilbert space, the bound is sharp: it provides a necessary and sufficient
condition that $W$ must satisfy in order to be attainable. 
Furthermore, the bound can be attained by carrying out
separate measurements on each particle.  This
completely solves the problem of estimating the state of a large
ensemble of spin 1/2 particles (qubits) in the same pure state. 

\item 
When $\rho(\theta)$ is a pure state belonging to a Hilbert space
of dimension $d$ larger then 2, then our bound on $W$ applies, but 
it is not sharp. 

\item 
When the unknown state is mixed and belongs to a 2 dimensional
Hilbert space, and if one restricts oneself to measurements that act
separately on each particle, then our bound applies and is sharp.

\item
When the unknown state is mixed and belongs to a 
Hilbert space of dimension $d>2$,
and if one restricts oneself to measurements that act
separately on each particle, then our bound applies but is not sharp.

\item
If the unknown state is mixed and one allows collective
measurements, then our bound is not necessarily safisfied.
\end{itemize}

This last point is surprising and points  to a
fundamental difference between measuring pure states and mixed
states. Indeed it is known that carrying out measurements on several
identical copies of the same pure state can generally be done better with 
collective measurements on the different
copies\cite{PeresWootters}\cite{MassarPopescu}. This is known
as `non-locality without entanglement'\cite{BDFMRSS}. The first
point shows that in the limit of a large number of copies, pure states
of spin 1/2 do not exhibit non-locality without entanglement. On the
other hand the last point shows that in the limit of a large number of
copies mixed states of spin 1/2 continue to exhibit non locality without
entanglement. 

To describe our bound on $W$, we first consider for simplicity the case of 
a pure state of spin 1/2 particles.  Suppose the unknown state is a spin
1/2 known to be in a pure state, and the state is known to be almost pointing 
in the $+z$ direction:

\begin{equation}
\left|\psi(\theta_1, \theta_2)\right\rangle \simeq 
\left|\uparrow_z\right\rangle + {1\over 2}(\theta_1 + i \theta_2)
\left|\downarrow_z\right\rangle
\label{psi}
\end{equation}
where we have written an expression valid to first order in
$\theta_1, \theta_2$. 
Suppose we carry out a measurement of the
operator $\sigma_x$. We obtain the outcome $\pm x$ with probability
$p(\pm x)=  ( 1 \pm \theta_1)/2$.  Thus the outcome of this measurement 
tells us about the value of $\theta_1$.
Similarly we can carry out a measurement of
$\sigma_y$.  We obtain the outcome $\pm y$ with probability
$p(\pm y)= (1 \pm \theta_2)/2$. The outcome of this measurement tells us about
$\theta_2$. But the measurements $\sigma_x$ and $\sigma_y$ are
incompatible, i.e., the operators do not commute and cannot be
measured simultaneously. Thus if one obtains knowledge about
$\theta_1$, it is at the expense of $\theta_2$. Indeed suppose one has
$N$ copies of the state $\psi$ and one measures $\sigma_x$ on $N_1$ copies
and $\sigma_y$ on $N_2 = N - N_1$ copies. 
Our estimator for $\theta_1$ is the fraction of $+x$ outcomes minus
the fraction of $-x$ outcomes. This estimator is unbiased.
The resulting uncertainty (at the point $\theta_1=\theta_2=0$)
about $\theta_1$ is then $\mbox{E}_\theta((\hat \theta_1 - \theta_1)^2) 
= { 1 \over N_1}$. Similarly we can estimate $\theta_2$ and the 
corresponding uncertainty is
$\mbox{E}_\theta((\hat \theta_2 - \theta_2)^2) = { 1 \over N_2}$. 
We can combine these two expressions in the following relation:

\begin{equation}
{1 \over \mbox{E}_\theta((\hat \theta_1 - \theta_1)^2 )}
+{ 1 \over \mbox{E}_\theta((\hat \theta_2 - \theta_2)^2) } =
{1 \over V^N_{11}} + {1 \over V^N_{22}} = N
\label{simple}
\end{equation}
which expresses in a compact form how we can
trade knowledge about $\theta_1$ for
knowledge about $\theta_2$. We shall show that it is impossible to do 
better than
precisely \eqref{simple} when one restricts attention to unbiased 
estimators based on
arbitrary measurements, 
and asymptotically not possible to do better with any estimator whatsover.

To generalize \eqref{simple}, we rewrite it in a
more abstract form, and state it as an inequality. We use polar coordinates to
parameterize the unknown state of the spin 1/2 particle:
$|\psi\rangle = \cos{\eta\over 2} |\uparrow\rangle +
\sin{\eta\over 2} e^{i\varphi}|\downarrow\rangle$. We introduce the tensor

\begin{equation}
H_{\eta\eta}=1\quad , \quad H_{\varphi\varphi}=\sin^2\eta\quad , \quad
H_{\eta\varphi}=0
\label{Ftwo}
\end{equation}
which is simply the Euclidean metric on the sphere. Then the bound 
\eqref{simple} can be reexpressed as

\begin{equation}
\Tr H^{-1} (V^N)^{-1} \le N
\label{simple2}
\end{equation}
where $V^N$ is the m.q.e.\ matrix defined in \eqref{V}. 

For mixed states belonging to a 2 dimensional Hilbert
space, \eqref{simple2} can be generalized as follows.
Let us suppose that the state $\rho(\theta)$ depends on three unknown 
parameters. 
Then we can parameterize it by $\rho(\theta) = 
{ 1 \over 2}( I + \sum_i\theta_i 
\sigma_i)$ where $ I$ is the 
identity matrix, $ \sigma_i$ are the Pauli matrices and the 3 parameters 
$ \theta_i$ obey $\|\theta\|^2 = \sum_i {\theta_i}^2 \le 1$. 
We now introduce the 
tensor 

\begin{equation}
H_{ij}(\theta) = \delta_{ij} + {{\theta_i \theta_j} \over 
{ 1 - \|\theta\|^2}}  
\label{F2d}
\end{equation}
which generalizes the tensor \eqref{Ftwo} to the case of mixed states.
Then, upon restricting oneself to separable measurements, 
we will show that the  
m.q.e.\ matrix $V^N$ must satisfy (exactly for unbiased estimators, 
and otherwise asymptotically)

\begin{equation}
\mbox{tr}\, H(\theta)^{-1} V^N(\theta)^{-1} \le N\ .
\label{const2d}
\end{equation}

As an application of these results, the minimum of the cost function 
(\ref{meancost}) in the case of spin 1/2 particles (for mixed states
restricting oneself to separable measurement) is 

\begin{equation}
\min\mbox{E}_\theta ({f}(\hat\theta,\theta)) = f_0(\theta) 
+ {\left( \Tr \sqrt{ H(\theta)^{-1/2} C(\theta) H(\theta)^{-1/2} } 
\right)^2 \over N} 
+ o(1/N)
\label{mincost}
\end{equation}
which is obtained simply by minimizing  
\eqref{meancost} subject to the constraints \eqref{simple2} or  
\eqref{const2d}.

We can compare \eqref{mincost} with the exact results which are known
in the case of 
covariant measurements on pure states of spin 1/2
particles \cite{Holevo}\cite{MassarPopescu}. In this problem one is given $N$
spin 1/2 particles polarized along the direction $\Omega$. The
$\Omega$  are uniformly distributed on the sphere. One wants to
devise a measurement and estimation strategy that minimize the
mean value of the cost function $\cos^2 {\omega/2}$, where $\omega$ is
the angle between the estimated direction $\hat \Omega$ and the true
direction $\Omega$. Expanding the cost function to second order in
$\omega$ (to obtain the quadratic cost matrix $C$) and averaging
\eqref{mincost} over the sphere, one finds

\begin{equation}
{\mbox E} (\cos^2 {\omega/2}) \geq 1 - {1\over N} +o({1\over N})
\end{equation}
which in the limit for large $N$ coincides with the results (exact for
all $N$) of \cite{Holevo}\cite{MassarPopescu}. 
If the directions $\Omega$ are not uniformly distributed, then 
\cite{Holevo}\cite{MassarPopescu} do not apply, but \eqref{mincost}
stays valid. However we cannot compare
our results with the recent analysis of covariant measurements on mixed
states\cite{mixedS} because we suppose
separability of the measurement, whereas \cite{mixedS} does not.

Equations \eqref{simple2} and \eqref{const2d} 
have a simple generalization to the
case of particles belonging to higher dimensional Hilbert spaces. But
in these cases these bounds are no longer sharp.

In order to appreciate the above results, 
we must recall some results from
classical statistical inference. This is the subject of the next section.

\section{Classical Cram\'er-Rao bound}

Consider a random variable $X$ with probability density $p(x,\theta)$.  
The connection with the quantum problem is that we
can view $p(x,\theta)$ as the probability density that a quantum measurement on
the system yields outcome $x$ given that the state was $\rho(\theta)$.
We take a random sample of size $N$ from the distribution and use it to
estimate the value of each parameter $\theta_i$. Call $\hat \theta_i^N$
the estimated value. The following results about the m.q.e.\ matrix of the
estimator are well known:

\begin{enumerate}

\item
Suppose that the estimator is unbiased, that is $\mbox{E}_\theta
(\hat \theta^N - \theta) =0$, where $\mbox{E}_\theta$ is the 
expectation value at 
fixed $\theta$, i.e., the integral $\int dx p(x|\theta)$.
Define its m.q.e.\ matrix $V^N(\theta)$ by

\begin{equation}
V^N_{ij}(\theta) =
\mbox{E}_\theta(
(\hat \theta_i^N - \theta_i) (\hat \theta_j^N - \theta_j)).
\end{equation}
Furthermore define the Fisher information matrix 
$I(\theta)$ by

\begin{eqnarray}
I_{ij}(\theta) &=& \mbox{E}_\theta( \partial_{\theta_i} \ln 
p(X|\theta)\partial_{\theta_j} \ln
p(X|\theta) )\nonumber\\
&=& \int dx {  \partial_{\theta_i}  p(x|\theta)\partial_{\theta_j} 
p(x|\theta) \over p(x|\theta) }\ .
\end{eqnarray}
Then for any $N$, the following inequalities, known as the Cram\'er-Rao
inequalities, hold\cite{Cramer}\cite{Helstrom}

\begin{equation}
V^N(\theta) \ge I(\theta)^{-1} / N
\label{cr1}
\end{equation}
or equivalently

\begin{eqnarray}
{V^N(\theta)}^{-1} \le N I(\theta),
\label{cr2}
\end{eqnarray}
the inequality meaning that the difference of the two sides is a
nonnegative matrix.

\item The hypothesis of unbiased estimators is very restrictive since
most estimators will be biased. Happily it is possible to relax 
this condition. Here are just two of the many results available:

\begin{enumerate}

\item First of all, if one is interested in averaging the mean cost
over possible values of $\theta$ with respect to a given prior distribution 
$\lambda(\theta)$, then there is a Bayesian
version of the Cram\'er-Rao inequality, the van Trees 
inequality \cite{vanTrees}\cite{GillLevit}. 
In the multivariate case, upon giving oneself a quadratic cost 
function determined by a matrix $C(\theta)$, one can derive the inequality

\begin{eqnarray}
\int d\theta \lambda (\theta)  
{\rm tr}\ C(\theta) V^N(\theta) \ge
{\int d\theta \lambda (\theta) {\rm tr}\  C(\theta) I^{-1}
(\theta) \over N} - {\alpha \over N^2}
\end{eqnarray}
where $\alpha$ is a positive number that depends on $C(\theta)$, $I(\theta)$, 
and $\lambda (\theta)$, but is independent of $N$.

\item 
The second approach makes no reference to any prior distribution for $\theta$,
but only holds in the limit $N$ tending to infinity and lays
a mild restriction on the estimators considered. 
Specifically, if the probability distribution of
$\sqrt{N} (\hat \theta^N - \theta) $ converges uniformly in $\theta$ towards 
a distribution depending continuously on $\theta$, say of a random vector $Z$, 
then the limiting scaled m.q.e.\ matrix $W(\theta)$
defined by $W_{ij}(\theta)=E_\theta(Z_i Z_j)$ obeys $W  \ge I^{-1}$.

\end{enumerate}

\item Furthermore in the
limit of arbitrarily large samples one can attain the Cram\'er-Rao
bound. This is proven by explicitly constructing an estimator that
attains the bound in the extended senses 2a) (apart from the $1/N^2$ term)
or 2b) just indicated: the maximum likelihood estimator (m.l.e.). 

\end{enumerate}

Modern statistical theory contains many other results having the
same flavour as point 2 above, namely that 
the Cram\'er-Rao bound holds in an approximate sense for
large $N$, without the restriction to biased estimators. Result 2a)
applies to a larger class of estimators than 2b), but only gives a result
on the average behaviour over different values of $\theta$. 
On the other hand combining results 3) and 2b) 
tells us that the maximum likelihood estimator is for large
$N$ an optimal estimator for each value of $\theta$ separately.
The reason why in 2b) additional regularity is demanded is because of
the phenomenon of super-efficiency (see \cite{superefficiency} for a recent 
discussion) whereby an estimator can have mean
quadratic error of smaller order than $1/N$ at isolated points.
Modern statistical theory (see again \cite{superefficiency} or
\cite{IbragimovHasminskii}) has concentrated on the more
difficult problem of obtaining non-Bayesian results
(i.e., pointwise rather than average) making much use of the technical
tool of `local asymptotic normality'. A major challenge in the quantum
case is to
obtain a result of type 2b) when this technique is definitely not available.

\section{Quantum  Cram\'er-Rao bound}

In this paper we show that similar results to 1, 2a, 2b, 3 can be obtained
when one must estimate the state of an
unknown quantum system $\rho(\theta)$ of which one possesses $N$ copies. 
This problem is most simply addressed, following \cite{BraunsteinCaves}, 
by decomposing it into a first (quantum)
step in which one carries out a measurement on $\rho^N = \rho \otimes
\dots \otimes \rho$ and a second (classical) step 
in which one uses the result of
the measurement to estimate the value of the parameters $\theta$. 

The most general way to describe the measurement is by a positive
operator-valued measurement (POVM) $M=(M_\xi)$ whose elements 
satisfy $M_\xi \ge 0$, $\sum_\xi M_\xi = I$.
(For simplicity we take the outcomes of the POVM to be discrete. 
The generalization to an arbitrary outcome space 
is just a question of translating into measure-theoretic language.)

Quantum mechanics tells us the probability to obtain outcome $\xi$
given state $\rho(\theta)$:

\begin{equation}
p(\xi|\theta) = \Tr \rho^N(\theta) M_\xi\ .
\end{equation}
  From the outcome $\xi$ of the measurement one must guess what are the
values of the $p$ parameters $\theta_i$. Call $\hat \theta^N$ the
estimated value of the parameter vector.
We want to obtain bounds on the m.q.e.\ matrix $V^N(\theta)$ 
of the estimator $\hat\theta^N$ when the true parameter value is 
$\theta$, thus $V_{ij}^N(\theta)=\mbox{E}_\theta
(\hat \theta_i^N - \theta_i) (\hat \theta_j^N - \theta_j)$.
To proceed we temporarily make the simplifying assumption that 
the estimators are unbiased,
$\mbox{E}_\theta  \hat\theta^N  = \theta$. 
Then we can apply the classical Cram\'er-Rao inequality to
the probability distribution $p(\xi|\theta)$ to obtain:

\begin{equation}
V^N \ge
I^N(M, \theta)^{-1}
\end{equation}
or

\begin{equation}
(V^N)^{-1} \le
I^{N}(M, \theta)
\end{equation}
where the Fisher information matrix $I^N$ for the measurement $M$
is defined by

\begin{eqnarray}
I_{ij}^N(M, \theta) &=& \sum_\xi {\partial_{i} p(\xi|\theta) 
\partial_{j} p(\xi|\theta) \over p(\xi|\theta)} \nonumber\\
&=&
\sum_\xi {
\Tr (\rho^N_{,i} M_\xi) \Tr (\rho^N_{,j} M_\xi) 
\over
\Tr (M_\xi \rho^N) }
\label{fisherXI}
\end{eqnarray}
with $\rho^N_{,i} = \partial_{\theta_i} \rho^N$.

These expressions suggest the following questions: 

\begin{enumerate}

\item
is there a simple bound for the m.q.e.\ $V^N$ of unbiased estimators 
$\hat\theta^N$, or equivalently for the Fisher information $I^N(M,\theta)$? 

\item
is the bound also valid for 
sufficiently well behaved but possibly biased
estimators---at least  in the limit of large $N$?

\item can this bound 
be attained---at least in the limit of a large number of copies
$N$? 

\end{enumerate}

Most of the work on this subject has been devoted to answering the
question 1). We now recall what is known about these questions.

Suppose first the parameter $\theta$ is one-dimensional, $p=1$. The
symmetric logarithmic derivative (s.l.d.)\ $\lambda_\theta$ of $\rho$
is the Hermitian matrix defined implicitly by

\begin{equation}
\rho_{,\theta} = {
\lambda_\theta \rho + \rho \lambda_\theta \over 2}.
\end{equation}
In a basis where $\rho$ is diagonal, 
$\rho = \sum_k p_k \left| k\right\rangle\left\langle k\right|$, this
can be inverted to yield

\begin{equation}
(\lambda_\theta )_{kl} = (\rho_{,\theta} )_{kl} {2 \over p_k + p_l}.
\end{equation}
Then we have the bound

\begin{equation}
I^N_{\theta \theta}(M, \theta) 
\le N \Tr \rho \lambda_\theta \lambda_\theta .
\label{crone}
\end{equation}
Furthermore it was suggested in
\cite{BarndorffNielsenGill} how to adapt the classical
m.l.e.\  so as to attain, in the limit of large $N$,
the bound (\ref{crone}).

In the multiparameter case the bound based on the s.l.d.\ can be
generalized in a natural way. Define the s.l.d.\ along direction
$\theta_i$ by

\begin{equation}
\rho_{,i} =  {
\lambda_{i} \rho + \rho \lambda_{i} \over 2},
\end{equation}
and Helstrom's quantum information matrix $H$ by
\begin{equation} H_{ij} = 
\Tr \rho{ \lambda_{i}  \lambda_{j}
+  \lambda_{j} \lambda_{i} \over 2}.
\label{sld}
\end{equation}
(This is the same matrix that was introduced for spin 1/2 particles for a 
particular choice of parameters in (\ref{Ftwo}) and (\ref{F2d})).
Then one can prove the bound \cite{Helstrom},

\begin{equation}
I^N(M, \theta) \le N H (\theta).
\label{Bsld}
\end{equation}
(This can be deduced directly from (\ref{crone}) 
as proven in \cite{BraunsteinCaves}. Indeed since (\ref{crone}) holds
for each path in parameter space, it implies the matrix equation 
(\ref{Bsld})).

However this bound is in general not achievable. Another bound has been
proposed based on an asymmetric logarithmic  derivative (a.l.d.)\cite{YuenLax}
which in some cases is better than (\ref{Bsld}). 
Holevo \cite{Holevo} has proposed yet another bound that is stronger then both
the s.l.d.\ and the a.l.d.\ bound, but this bound is not explicit: it
requires a further minimization.
As far as we know no general achievable bound is known in the
multiparameter case. 

The difficulty in obtaining a
simple bound in the multiparameter case
is that there are many inequivalent ways in which one can
minimize the m.q.e.\  matrix $V_{ij}^N$. That is, in order to build a good
estimator one must make a choice of what one wants to estimate, and
according to this choice the measurement strategy followed will be
different. Hence a bound in the form of a matrix inequality like
(\ref{Bsld}) cannot be expected to be tight.

\section{Results}\label{Results}

In this paper we obtain answers to the three questions raised above in 
the multiparameter case. Our results are summarized in this section.

We first discuss point 1), that is bounds on
the Fisher information. We shall show the following:

\bigskip

\noindent {\bf Theorem I:\/}
When $\rho(\theta)=|\psi(\theta)\rangle\langle\psi(\theta)|$ is a pure
state, then 
the Fisher information
$ I^N(M,\theta)$ defined in (\ref{fisherXI})
must satisfy the following relation

\begin{equation}
\Tr H^{-1}(\theta) I^N(M,\theta)\  \le \ (d-1) N
\label{central}
\end{equation}
where $H^{-1}$ is the inverse of the quantum information matrix
defined in (\ref{sld}) and $d$ is the dimension of the Hilbert space 
to which $\rho(\theta)$ belongs. 
Note that this inequality (\ref{central}) is
invariant under change of parameterization $\theta \to \theta'(\theta)$.

\bigskip

This result immediately gives an inequality for the mean quadratic
error matrix 
of {\it unbiased\/} estimators $\hat\theta^N$
by invoking the classical Cram\'er-Rao inequality in order to replace
$I^N(M,\theta)$ by the inverse of the m.q.e.\ $V^N(\theta)$:

\begin{equation}
 \Tr H^{-1}(\theta) (N{ V}^N(\theta))^{-1} \ \le \  d-1\ .
\label{central2}
\end{equation} 

\bigskip

\noindent {\bf Theorem II:\/}  
When $\rho(\theta)$ is a mixed
state, and if the measurement $M$ consists of separate
measurements on each particle, then 
the Fisher information also satisfies \eqref{central}.
Hence for separable measurements on a mixed state, the m.q.e.\ matrix
of an unbiased estimator satisfies \eqref{central2}.

\bigskip

\noindent {\bf Theorem III (non additivity of optimal Fisher information):\/}
In the case of mixed states, it is in general possible to devise a
collective measurement for which the Fisher information does not
satisfy the inequality (\ref{central}).

\bigskip 

The second part of the paper 
consists in proving that the
constraint (\ref{central2}) also holds for biased estimators 
under suitable additional
conditions. We give two forms of this generalized form of
(\ref{central2}) corresponding to the two forms 2a) and 2b)
of the generalized
classical Cram\'er-Rao inequality.

Consider $N$ copies of a state $\rho(\theta)$. If $\rho$ is pure we
can make either collective or separable measurements. If $\rho$ is
mixed we restrict ourselves to separable measurements  (since Theorem
III shows that in this case collective measurements can beat
(\ref{central})).
Based on the outcome of the measurement we estimate the value of the
parameter vector $\theta$. Call $\hat \theta$ the estimator,
and denote by $V^N=V^N(\theta)$ its m.q.e.\ matrix when the true value
of the parameter is $\theta$.

We shall prove the following generalization of result of type 2b)
concerning the behaviour of the mean quadratic error matrix as $N$ tends to
infinity: 

\bigskip

\noindent {\bf Theorem IV:\/}
Suppose that the scaled m.q.e.\ $N V^N(\theta)$ has the limit 
$W(\theta)$ as $N\to\infty$.
Suppose that the convergence is uniform 
in $\theta$ and that $W$ is continuous at the point
$\theta=\theta^0$. Furthermore we suppose that $H$ and its derivatives
are bounded in a neighbourhood of this point.
Then we shall prove in section \ref{optimality} that
$W(\theta^0)$ must satisfy

\begin{equation}
 \Tr H^{-1}(\theta^0) W^{-1}(\theta^0) \leq (d-1).
\label{central3}
\end{equation} 

\bigskip

This result gives a bound on the mean value of a quadratic cost
function $C$ as $N$ tends to infinity. Indeed
using a Lagrange multiplier to impose the condition (\ref{central3}), 
the minimum cost is readily found to be

\begin{equation}
\lim_{N\to \infty} N \Tr C(\theta^0) V^N(\theta^0) \ge 
\left( \Tr \sqrt{ H^{ -{1\over 2}}(\theta^0) C(\theta^0)
 H^{ -{1\over 2}}(\theta^0)}  \right)^2 .
\label{CVNopt}
\end{equation}

In terms of a cost function, it is also possible to prove a Bayesian
version of the Cram\'er-Rao inequality which is the analogue of the
classical result 2a):

\bigskip

\noindent {\bf Theorem V:\/}
Suppose that one is given a quadratic cost function $C(\theta)$ and 
a prior distribution $\lambda(\theta)$ for the
parameters $\theta$. If $C$, $\lambda$ and $H$ are sufficiently
smooth functions of $\theta$ (continuity of the first derivatives
is sufficient), while $\lambda$ is zero outside a compact region with
smooth boundary, then

\begin{equation}
\int d\theta \lambda(\theta) \Tr C(\theta) {V}^N(\theta)
\geq {1 \over N}
\int d\theta \lambda(\theta)\Tr \left( 
\sqrt{ H^{ -{1\over 2}}(\theta) C(\theta)
 H^{ -{1\over 2}}(\theta)}  \right )^2
- {\alpha \over N^2}
\label{central3b}
\end{equation}
where $\alpha$ is a constant independent of $N$ but which depends on 
$C$, $\lambda$ and $H$.

\bigskip

Theorems I, II, IV and V put bounds on the m.q.e.\ matrix of an estimator of
an unknown state $\rho(\theta)$ (for mixed states, under the restriction that
the measurement is separable).
The third part of this article is devoted to showing that in the case
of spin 1/2 systems ($d=2$)
these bounds can be attained.
We first show that at any point
$\theta^0$ we can attain equality in \eqref{central}.

\bigskip

\noindent {\bf Theorem VI:\/}
Suppose one has
$N$ spin 1/2 particles in an unknown (possibly mixed) state
$\rho(\theta)$. Fix any point $\theta^0$. 
Give yourself a matrix $G^0$ satisfying
$\Tr H^{ -1}(\theta^0) G^0\le 1$.
We call $G^0$ the target scaled information matrix.
Then there exists a measurement $M^{\theta^0}$ (depending
on the choice of $\theta^0$) acting on
each spin separately such that
$I^N(M^{\theta^0}, \theta^0) = N G^0$. This measurement is described
in detail in section \ref{2dtheta0}.

\bigskip

For large $N$ we can  
also approximately attain equality at all points $\theta$ simultaneously:

\bigskip

\noindent {\bf Theorem VII: \/}
Suppose one has $N$ spin 1/2 particles in an
unknown pure state $|\psi(\theta)\rangle$ 
or suppose that one has $N$ spin 1/2 particles in an
unknown mixed state $\rho(\theta)$. 
In the latter case we also require that the
state never be pure, i.e. $\Tr \rho(\theta)^2 < 1$ for all $\theta$.

Give oneself a smooth positive matrix function $G(\theta)$ satisfying
$\Tr H^{ -1}(\theta) G (\theta)\le 1$ for all $\theta$, the target scaled
information for each possible value of $\theta$. 
Define the corresponding target scaled m.q.e.\ matrix
$W(\theta)=G(\theta)^{-1}$. Suppose that $W(\theta)$ is non singular
(i.e. $G(\theta)$ never has a zero eigenvalue). 

Then there exists a measurement $M$ acting on
each spin separately, and a corresponding estimator 
$\hat\theta$, whose m.q.e.\ matrix $V^N(\theta)$ satisfies

\begin{equation}
V^N(\theta)= 
\frac {W(\theta)} N + o(1/N)
\label{eeref}
\end{equation}
for all values of $\theta$ simultaneously.
For this estimation strategy
$\sqrt{N} (\hat \theta - \theta) $ converges in distribution
towards $N(0,W)$, the normal distribution with mean zero and
covariance $W$. The measurement $M$ and estimation 
strategy is described in detail in
section \ref{sub2}.

It is interesting to note that the measurement strategy which
satisfies \eqref{eeref} is an adaptive one. That is one first carries
out a measurement on a small fraction of the particles. This gives a
preliminary estimate of the quantum state which allows a fine tuning
of the measurements that are carried out on the remaining
particles. This is to be contrasted to previously proposed state
estimation strategies in the case of finite dimensional Hilbert spaces 
\cite{weigert}\cite{Leonhardt} 
in which the same measurement is carried out on all the 
particles. The necessity of an adaptive measurement strategy if one
wants to minimise the m.q.e.\ was pointed out in \cite{BarndorffNielsenGill}.

When the unknown state belongs to a Hilbert space of dimension $d>2$,
then the bound \eqref{central} cannot be attained in general. Indeed
we shall show in section \ref{Comp}
that for $d>2$, neither \eqref{Bsld} nor \eqref{central} implies the other.

\section{New quantum Cram\'er-Rao inequality}\label{Central}

In this section we prove Theorems I, II, III. That is we prove 
(\ref{central}) for general measurements in the case of
pure states and for separate measurements on each particle in the case
of mixed states.

\subsection{Preliminary results}\label{VA}

The first step in proving (\ref{central}) is to show that one can restrict
oneself to POVM's whose elements
are proportional to one dimensional projectors.
Indeed any POVM can always be refined to yield a POVM whose elements are 
proportional to one
dimensional projectors. We call such a measurement {\it exhaustive}. 
This yields a refined probability
distribution $(p(\xi, \theta))$. It is well known that under such refining 
of the probability distribution, the Fisher information can only 
increase\cite{GeneralCramerRao}.

The second step in proving (\ref{central}) consists in increasing
the number of parameters. Suppose that $\rho(\theta)$ depends on $p$
parameters $\theta_i$, $i=1,\dots,p$. If $\rho =|\psi(\theta)\rangle
\langle \psi(\theta)|$ is a pure state, then $p\le 2d-2$ (since
$|\psi(\theta)\rangle$ is normalized and defined up to a phase).
If $\rho$ is a mixed state, then Hermiticity and the condition
$\Tr \rho =1$ impose that $p\le d^2-1$. 
Suppose that $p < \nu$ is less then the maximum number of possible
parameters ($\nu=2d-2$ or $\nu=d^2-1$ according to whether the state is pure
or mixed). Then 
one can always increase the number of parameters up to
the maximum. Indeed let us suppose that to the $p$ parameters, one
adds independent parameters
$\theta_{i'}$, $i'=p+1,\dots,\nu$.
We can now consider the quantum information matrix $\tilde H$, and
Fisher information matrix $\tilde I$, for the completed set of parameters. 
We shall show below that 

\begin{equation}
\Tr H^{-1}(\theta) I^N(M, \theta)
\le  
\Tr {\tilde H}^{-1} (\theta) {\tilde I^N}(M,\theta).\label{reduc}
\end{equation}
Therefore it will be sufficient to prove \eqref{central} in the
case when there are $\nu$ parameters.

To prove \eqref{reduc}, fix a particular point $\theta^0$. At this
point we have the derivative $\rho_{,i}$ and s.l.d.\ $\lambda_i$ of
$\rho$ for $i=1,\dots,p$. Introduce a set of Hermitian matrices $\lambda_{i'}$ 
with $\Tr \rho(\theta^0)\lambda_{i'}=0$, for $i' = p+1,\dots,\nu$, such that

\begin{equation}
\Tr \rho(\theta^0 ){\lambda_i \lambda_{i'} + \lambda_{i'} \lambda_i
\over 2} = 0 \quad , \quad i=1,\dots,p \quad , \quad i'=p+1,\dots,\nu.
\label{rll}
\end{equation}
This is always possible because we can view 
\eqref{rll} as a scalar product between $\lambda_i$ and
$\lambda_{i'}$ and a Gram-Schmidt orthogonalization procedure will
then yield the matrices $\lambda_{i'}$. Now define matrices
$\rho_{,i'}$ by   $\rho_{,i'} = 
( \rho(\theta^0 )  \lambda_{i'} +  \lambda_{i'} \rho(\theta^0 ) )/2$
and define additional parameters $\theta_{i'}$ satisfying, 
at $\theta^0$, $\partial_{\theta_{i'}} \rho = \rho_{,i'}$.
The point of this construction is that because of \eqref{rll}, the
quantum information matrix $\tilde H$ is block diagonal with the
first block equal to $H$. Let $\tilde I(M)$ 
be the Fisher information matrix for the enlarged 
set of parameters (but the same measurement).  Then
$\Tr {\tilde H}^{-1} \tilde I (M)
=\Tr ({\tilde H}^{-1})_{11} {\tilde I}_{11}(M) + 
 \Tr ({\tilde H}^{-1})_{22} {\tilde I}_{22}(M)$ 
where the indices $11$ and $22$ denote the blocks of these
matrices corresponding to the original and the new parameters. But
both terms are nonnegative since all matrices involved are
nonnegative, and $({\tilde H}^{-1})_{11}=H^{-1}$, 
so we obtain \eqref{reduc} at $\theta_0$ and for the particular 
parameters just introduced. But since the right hand side of \eqref{reduc}
is invariant under reparametrization, it is true for any parametrization,
and at any $\theta$.

\subsection{Pure states}\label{VB}

To proceed we shall consider a 
POVM whose elements are proportional to one dimensional projectors
and calculate explicitly the left hand side of
\eqref{central} in the case where the number of parameters is
the maximum  $p = 2d-2$ in a basis where $H$ is diagonal.

We fix a point $\theta^0$. At this point we chose a basis
such that

\begin{equation}
\rho(\theta^0) = |1 \rangle \langle 1|\ .
\end{equation}
Hence the density matrix of the $N$ copies is

\begin{equation}
\rho^N = |1\rangle\langle 1| \otimes \ldots\otimes |1\rangle\langle 1|\ .
\end{equation}

Consider now the $2d-2$ Hermitian operators

\begin{eqnarray}
\rho_{,k+} &=& \left| 1\right\rangle\left\langle k\right| + \left| 
k\right\rangle\left\langle 1\right| \quad , \quad 1<k\leq d \ ,\nonumber\\
\rho_{,k-} &=& i\left| 1\right\rangle\left\langle k\right| -i \left| 
k\right\rangle\left\langle 1\right| \quad , \quad 1<k \leq d \ .
\label{cshm1}
\end{eqnarray}
We choose a parameterisation such that in the vicinity of
$\theta^0$, it has the form $\rho = \rho(\theta^0) + \sum_{k,\pm}
(\theta_{k\pm} - \theta_{k\pm}^0) \rho_{,k\pm}$ with the unknown parameters
$\theta_{k\pm}$, $k=2,\dots,d$. With this parametrisation the 
derivatives of $\rho^N$ are 

\begin{equation}
\rho^N_{,k\pm} = \rho_{,k\pm}\otimes \rho \dots\otimes \rho+\dots
+\rho \otimes\dots\otimes \rho_{,k\pm}\ .
\end{equation}

One then calculates the s.l.d.\ of $\rho$ and hence the 
quantum information matrix $H$. One verifies that in this basis $H$ is
diagonal:

\begin{eqnarray}
H_{k\pm, k'\pm'} &=& {4 } \delta_{kk'} 
\delta_{\pm \pm'} \ .
\end{eqnarray}

Consider any POVM whose elements are proportional to one dimensional
projectors

\begin{equation}
M_\xi = \left| \psi_\xi\right\rangle\left\langle \psi_\xi\right|
\quad , \quad
\left| \psi_\xi\right\rangle = \sum_{k_1=1}^d \ldots\sum_{k_N=1}^d
a_{\xi k_1\ldots k_N} \left|k_1\ldots k_N\right\rangle\ .\end{equation}
The completeness relation $\sum_\xi M_\xi = I$ takes the form

\begin{equation}
\sum_\xi a_{\xi k_1\ldots k_N}^* a_{\xi k_1' \ldots k_N'}=
\delta_{k_1 k_1'} \ldots \delta_{k_N k_N'} \ .
\end{equation}

To proceed we need the following formulae:

\begin{equation}
\Tr \rho(\theta^0)M_{\xi} = 
|a_{\xi 1\ldots1}|^2 
\end{equation}
and

\begin{equation}
\Tr \rho(\theta^0)_{,k+}M_\xi = \sum_{p=1}^N 
(a_{\xi 1 \ldots 1}^* a_{\xi 1 \ldots k_p=k \ldots 1}
+ a_{\xi 1 \ldots k_p=k \ldots 1}^* a_{\xi 1 \ldots 1})
\end{equation}
and similarly for $\Tr \rho(\theta^0)_{,k-}M_\xi$.
Thus we obtain

\begin{equation}
(\Tr \rho(\theta^0)_{,k+}M_\xi)^2 
+ (\Tr \rho(\theta^0)_{,k-}M_\xi)^2 
= \sum_{p=1}^N 4
|a_{\xi 1 \ldots 1}|^2 |a_{\xi 1\ldots k_p=k \ldots 1}|^2 \ .
\end{equation}

Putting everything together yields

\begin{eqnarray}
\Tr H^{ -1} I(M)
&=& \sum_\xi {1 \over \Tr \rho(\theta^0)M_\xi }{1 \over 4}
\sum_{k=2}^d \sum_\pm (\Tr \rho(\theta^0)_{,k+}M_\xi)^2 
+ (\Tr \rho(\theta^0)_{,k-}M_\xi)^2 \nonumber\\
&=& \sum_{k=2}^d \sum_{p=1}^N \sum_\xi |a_{\xi
1 \ldots k_p=k \ldots 1}|^2\nonumber\\
&=& N(d-1)
\end{eqnarray}
which proves that equality holds in \eqref{central}
for arbitrary exhaustive measurements in the case of pure states.

\subsection{One mixed state}\label{onemixedstate}

Deriving \eqref{central} for  
mixed states is more complicated than for pure,  and we shall
proceed in two steps. First we shall consider the case of one mixed
state ($N=1$) and show that equality in (\ref{central}) holds in this
case for arbitrary exhaustive measurements. Then we shall consider the 
case of an arbitrary number $N$ of mixed states.

We first diagonalize $\rho$ at a point $\theta^0$: 
$\rho(\theta^0) = \sum_{k=1}^d p_k \left| k\right\rangle\left\langle k\right|$.
We now introduce the following complete
set of Hermitian traceless matrices:

\begin{eqnarray}
\rho_{,kl+} &=& \left| k\right\rangle\left\langle l\right| + \left| 
l\right\rangle\left\langle k\right| \quad , \quad k<l \ ,\nonumber\\
\rho_{,kl-} &=& i\left| k\right\rangle\left\langle l\right| -i \left| 
l\right\rangle\left\langle k\right| \quad , \quad k<l \ ,\nonumber\\
\rho_{,m} &= & \sum_{k = 1}^d c_{mk}
\left| k\right\rangle\left\langle k\right| 
\quad , \quad m=1,\ldots ,d-1
\label{cshm}
\end{eqnarray}
where  the coefficients $c_{mk}$ obey

\begin{eqnarray}
\sum_k c_{mk} &=&0 \ ,\nonumber\\
\sum_k {1 \over p_k} c_{m'k}c_{mk} &=& \delta_{m'm}.
\label{cm}
\end{eqnarray}
Let us denote 
the matrices $\rho_{,kl\pm}$ and $ \rho_{,m}$ collectively as
$\rho_{,i}$. (They
constitute a set of generators
of $\mbox{su}(d)$).

We choose a parameterization such that in the vicinity of
$\theta^0$, it has the form $\rho = \rho(\theta^0) + \sum_i (\theta_i
- \theta^0_i) \rho_{,i}$.
One then calculates the s.l.d.\ of $\rho$ and from this the
quantum information matrix $H$. 
One verifies that in this basis $H$ is diagonal:

\begin{eqnarray}
H_{kl\pm, k'l'\pm'} &=& {4 \over p_k + p_l} \delta_{kk'} \delta_{ll'}
\delta_{\pm \pm'} \ ,\nonumber\\
H_{kl\pm,m} &=& 0 \ ,\nonumber\\
H_{m,m'} &=& \delta_{m'm}\ .
\end{eqnarray}

Consider any POVM whose elements are proportional to one dimensional
projectors

\begin{eqnarray}
M_\xi &=& \left| \psi_\xi\right\rangle\left\langle \psi_\xi\right|\ , 
\nonumber\\
\left| \psi_\xi\right\rangle &=& \sum_k a_{\xi k} \left|
  k\right\rangle\ .
\end{eqnarray}

The l.h.s.\ of (\ref{central}) can now be written as

\begin{equation}
\Tr H^{ -1} I(M)
= \sum_\xi {1 \over \left\langle \psi_\xi| \rho |
\psi_\xi\right\rangle }
\left( \sum_{k<l} \sum_\pm
{p_k + p_l \over 4} \left\langle\psi_\xi | \rho_{,kl\pm} | 
\psi_\xi\right\rangle^2
+ \sum_m  \left\langle\psi_\xi | \rho_{,m} | \psi_\xi\right\rangle^2
\right).
\label{First}
\end{equation}
Using the following expressions

\begin{eqnarray}
\left\langle\psi_\xi | \rho_{,m} | \psi_\xi\right\rangle &=& \sum_k |a_{\xi 
k}|^2 c_{mk}\ ,
\nonumber\\
 \left\langle\psi_\xi | \rho_{,kl+} | \psi_\xi\right\rangle^2
+  \left\langle\psi_\xi | \rho_{,kl-} | \psi_\xi\right\rangle^2 &=& 4 | 
a_{\xi k}|^2| a_{\xi l}|^2
\end{eqnarray}
one obtains

\begin{eqnarray}
\Tr H^{ -1} I(M)
&=& \sum_\xi {1 \over \left\langle \psi_\xi| \rho | \psi_\xi\right\rangle }
\left( \sum_{k<l} (p_k + p_l) | a_{\xi k}|^2| a_{\xi l}|^2
+ \sum_m (\sum_k  |a_{\xi k}|^2 c_{mk} )^2 \right)\nonumber\\
&=& \sum_\xi {1 \over \left\langle \psi_\xi| \rho | \psi_\xi\right\rangle }
\left( \sum_{k\neq l} p_k | a_{\xi k}|^2| a_{\xi l}|^2
+ \sum_k \sum_l   |a_{\xi k}|^2 |a_{\xi l}|^2 \sum_m c_{mk} c_{ml}  \right).
\label{fint}
\end{eqnarray}
We now use the following relation

\begin{equation}
\sum_m  c_{mk} c_{ml} = \delta_{kl} p_k - p_k p_l
\label{rel}\end{equation} which is derived  from (\ref{cm})
as follows: define $v_{mk}= c_{mk} / \sqrt{p_k}$ ($m=1,\ldots ,d-1$) and
$v_{dk} = \sqrt{p_k}$. Then (\ref{cm}) can be rewritten as
$\sum_k v_{mk} v_{m'k} = \delta_{mm'}$. The vectors $v_{mk}$ therefore
are a complete orthonormal basis of $R^d$, hence they obey $\sum_m
v_{mk} v_{mk'} = \delta_{kk'}$. Reexpressing in terms of $c_{mk}$
yields (\ref{rel}). Inserting it in (\ref{fint}) we obtain

\begin{eqnarray}
\Tr H^{-1} I(M)
&=&  \sum_\xi {1 \over \left\langle \psi_\xi| \rho | \psi_\xi\right\rangle }
\left( \sum_k \sum_l p_k (1 - p_l)|a_{\xi k}|^2 |a_{\xi
l}|^2\right)\nonumber\\
&=& \sum_k ( 1- p_k) \sum_\xi |a_{\xi k}|^2 
=\sum_\xi \Tr (I - \rho ) M_\xi
\nonumber\\
&=& d-1 
\label{Last}
\end{eqnarray}
as announced.

Note that this has demonstrated that equality holds in 
(\ref{central}) whenever
$N=1$, $p=d^2-1$, and the POVM is exhaustive. It follows 
from the classical properties of the Fisher information 
that equality also
holds for arbitrary $N$ whenever the POVM can be considered as a 
sequence of $N$ separate exhaustive measurements on each copy of the
system. It also  holds if the $n$'th measurement is chosen at random
depending on the outcomes of the previous measurements.

\subsection{Separable measurements on N mixed states}\label{Nmixedstates}

We shall now prove that if 
we possess $N$ identical mixed states of spin 1/2 particles, and carry 
out separable measurements, then

\begin{equation}
\Tr H^{-1} I(M) \le N (d-1).
\label{FIN}
\end{equation}
We recall that a separable measurement is one that can be carried out
sequentially on separate particles, where the measurement on one particle
at any stage (and indeed which particle to measure: one is allowed to measure 
particles several times) can depend arbitrarily
on the outcomes so far, see \cite{PeresWootters} 
for a discussion. It is therefore more general than the case
considered at the end of the previous subsection where the measurement 
on the $n$th particle could only depend on the measurements carried out 
on the  $n-1$ previous particles.

If a POVM is separable, then its 
elements $M_\xi$ can be decomposed into a sum of terms
proportional to projectors onto unentangled states

\begin{eqnarray}
M_\xi &=& \sum_i |\psi_{\xi i}\rangle \langle \psi_{\xi i} | \ ,
\nonumber\\
|\psi_{\xi i}\rangle &=& |\psi_{\xi i}^1\rangle
\otimes \ldots \otimes
 |\psi_{\xi i}^N\rangle \ .
\end{eqnarray}
We call  measurements having such a representation {\it nonentangled}.
(Note that there exist nonentangled POVMs which are not
separable\cite{BDFMRSS}). 

By refining a separable measurement (which increases the Fisher
information) one can restrict oneself to measurements whose POVM
elements are proportional to projectors onto product states

\begin{eqnarray}
M_\xi =  |\psi_{\xi }\rangle \langle \psi_{\xi } | =
|\psi_{\xi }^1\rangle \langle \psi_{\xi }^1 |
\otimes \ldots \otimes
 |\psi_{\xi}^N\rangle \langle \psi_{\xi }^N |\ .
\label{sep}
\end{eqnarray}

We now evaluate the l.h.s.\ of (\ref{FIN}) for measurements of the form
(\ref{sep}). First recall that the 
$N$ unknown states have the form

\begin{equation}
\rho^N = \rho \otimes \ldots \otimes \rho
=\sum_{k_1=1}^d
\cdots\sum_{k_N=1}^d p_{k_1} ... p_{k_N}
\left|k_1...k_N\right\rangle\left\langle k_1...k_N\right|
\end{equation}
and the derivatives of $\rho^N$ have the form

\begin{equation}
\rho^N_{,i} = \rho_{,i}\otimes \rho \ldots \otimes \rho+\ldots
+\rho \otimes...\otimes \rho_{,i}
=\sum_{p=1}^N \rho \otimes \ldots\rho_{,i} \ldots \otimes \rho \end{equation} 
where in the second rewriting it is understood that 
$\rho_{,i}$ is at the $p$'th position in the product.

Using the product form of measurement (\ref{sep}), one finds that

\begin{eqnarray}
\langle \psi_\xi | \rho^N |\psi_\xi \rangle &=&
\langle \psi_\xi ^1| \rho|\psi_\xi^1\rangle ...
\langle \psi_\xi ^N| \rho|\psi_\xi^N\rangle
\nonumber\\
\langle \psi_\xi | \rho^N_{,i} |\psi_\xi \rangle &=&\sum_{p=1}^N
\langle \psi_\xi ^1| \rho|\psi_\xi^1\rangle ...
\langle \psi_\xi ^p| \rho_{,i}|\psi_\xi^p\rangle ...
\langle \psi_\xi ^N| \rho|\psi_\xi^N\rangle
\end{eqnarray}
Inserting these expressions into the Fisher information matrix one finds

\begin{eqnarray}
I_{ij}(M)
&=& \sum_\xi { \langle \psi_\xi | \rho^N_{,i} |\psi_\xi \rangle
\langle \psi_\xi | \rho^N_{,j} |\psi_\xi \rangle \over
\langle \psi_\xi | \rho^N |\psi_\xi \rangle }
\nonumber\\
&=&\sum_\xi\sum_{p \neq p'}
\langle \psi_\xi ^1| \rho|\psi_\xi^1\rangle ...
\langle \psi_\xi ^p| \rho_{,i}|\psi_\xi^p\rangle ...
\langle \psi_\xi ^{p'}| \rho_{,j}|\psi_\xi^{p'}\rangle ...
\langle \psi_\xi ^N| \rho|\psi_\xi^N\rangle
\nonumber\\
& &+\sum_\xi\sum_{p}
\langle \psi_\xi ^1| \rho|\psi_\xi^1\rangle ...
{\langle \psi_\xi ^p| \rho_{,i}|\psi_\xi^p\rangle 
\langle \psi_\xi ^p| \rho_{,j}|\psi_\xi^p\rangle \over
\langle \psi_\xi ^p| \rho |\psi_\xi^p\rangle}
...
\langle \psi_\xi ^N| \rho|\psi_\xi^N\rangle
\nonumber\\
&=&\sum_\xi\sum_{p}
\langle \psi_\xi ^1| \rho|\psi_\xi^1\rangle ...
{\langle \psi_\xi ^p| \rho_{,i}|\psi_\xi^p\rangle 
\langle \psi_\xi ^p| \rho_{,j}|\psi_\xi^p\rangle \over
\langle \psi_\xi ^p| \rho |\psi_\xi^p\rangle}
...
\langle \psi_\xi ^N| \rho|\psi_\xi^N\rangle,
\label{two}
\end{eqnarray}
where we have used the fact that 
the first term in the second equality vanishes. Indeed it is equal to
\begin{eqnarray}
\sum_\xi\sum_{p \neq p'}
\langle \psi_\xi |\rho \otimes \ldots \otimes \rho_{,i}\otimes \ldots \otimes
\rho_{,j}\otimes \ldots \otimes\rho |\psi_\xi\rangle .
\label{four}
\end{eqnarray}
The sum over $\xi$ 
can be carried out in (\ref{four}) to yield the identity matrix and
the resulting trace vanishes because 
$\Tr \rho \otimes \ldots \otimes  \rho_{,i}\otimes \ldots \otimes
\rho_{,j}\otimes \ldots \otimes\rho =0$.

We now insert (\ref{two}) into $\Tr H^{-1} I(M)$.
All the operations from 
(\ref{First}) to (\ref{Last}) can be carried out exactly as in the previous
subsection, and one arrives at the expression

\begin{eqnarray}
\Tr H^{-1} I(M)
&=& \sum_{p}\sum_\xi \langle \psi_\xi |
\rho \otimes \ldots \otimes (I - \rho )\otimes \ldots
\otimes \rho |\psi_\xi\rangle
\nonumber\\
&=& N(d-1)
\end{eqnarray}
which is the sought for relation.

\subsection{Inequality for more then one mixed state}

We now provide a counterexample showing that if one carries out a
collective measurement on $N>1$ mixed states one can violate
(\ref{central}).
We take $N=2$, and suppose the unknown states belong to a 
$2$ dimensional Hilbert
space. $\rho(\theta) = {I \over 2} + \sum_i \theta_i \sigma_i$.
We take as reference point $\theta_i=0$ corresponding to 
$\rho = {I\over 2}$.
At this point $H_{ij} (\theta_i=0) = \delta_{ij}$.

We consider as measurement on the two copies the following POVM
\begin{eqnarray} M&=& \left\{\quad
{1\over 2} |\uparrow_x\uparrow_x\rangle \langle \uparrow_x\uparrow_x |
\quad , \quad
{1\over 2}
|\downarrow_x\downarrow_x\rangle\langle\downarrow_x\downarrow_x|
\quad , \quad
{1\over 2} |\uparrow_y\uparrow_y\rangle \langle \uparrow_y\uparrow_y
|
\quad , \quad
{1\over 2}
|\downarrow_y\downarrow_y\rangle\langle\downarrow_y\downarrow_y|
\quad ,
\right.\nonumber\\ & & \left.
{1\over 2} |\uparrow_z\uparrow_z\rangle \langle \uparrow_z\uparrow_z
|
\quad , \quad
{1\over 2}
|\downarrow_z\downarrow_z\rangle\langle\downarrow_z\downarrow_z|
\quad , \quad
{1\over 2} |\uparrow_z\downarrow_z - \downarrow_z\uparrow_z
\rangle \langle \uparrow_z\downarrow_z -
\downarrow_z\uparrow_z|\quad\right\} \ .
\end{eqnarray}
This POVM cannot be realized by separate measurements on each particle
since  the last term projects onto an entangled state.

For this POVM one calculates that
$I_{ij}(M, \theta_i=0) = \delta_{ij}$. 
Hence the left hand side of (\ref{central}) evaluates to
$\sum_{ij}H_{ij}^{-1} (\theta_i=0) I_{ij}(M, \theta_i=0) =3 > N(d-1) =2$.
This shows that the optimal Fisher information is non additive.

\subsection{Comparison with other Quantum Cram\'er-Rao bounds}\label{Comp}

An important question raised by the bound (\ref{central}) is how
it compares to other quantum Cram\'er-Rao bounds obtained in the
literature. In this respect, our
most important result is that (\ref{central}) is both a
necessary and sufficient condition that $I(M, \theta)$ must
satisfy when the dimensionality of the system $d$ equals $2$ and the
state is pure. This will be proven and discussed in detail in section \ref{2dim}. 

When $d>2$ (\ref{central}) is not a sufficient condition that 
$I(M, \theta)$ must satisfy. To see this let us compare
(\ref{central}) with the bound derived by Helstrom based on the
s.l.d.  This bound is the matrix inequality
$I^N(M,\theta) \le N H(\theta)$, see (\ref{Bsld}).
The comparison is most easily carried out by defining the matrix
$F = {1 \over N} H^{-{1\over 2}} I^N H^{-{1\over 2}} = \sum_{i=1}^p
\gamma_i f_i\otimes f_i$ where $\gamma_i$ are the eigenvalues of $F$
and $f_i$ its eigenvectors. Helstrom's bound can be reexpressed as
$\gamma_i\le 1$ for all $i$, 
whereas the bound \eqref{central} states that 
$\sum_i \gamma_i \le d-1$.
  From these expressions it results that the bound  \eqref{central} is
better then Helstrom's bound for $d=2$. For $d>2$ and $p \le
d-1$ Helstrom's bound is better then \eqref{central} as is seen by
summing the inequalities $\gamma_i\le 1$ to obtain $\sum_i
\gamma_i \le p$. For $p > d-1$,  neither Helstrom's bound nor the bound
(\ref{central}) are better than the other. 

Yuen and Lax\cite{YuenLax}  have proposed another matrix bound based on an
asymmetric logarithmic derivative (a.l.d.). This bound is known to be 
worse then the bound based on the s.l.d.\ in the
case of one parameter, but it can be better, for some loss functions,
in the case of two or more parameters. 
We have however not been able to make a detailed 
comparison between the bound based on the a.l.d.\ and (\ref{central}).

Although when $d>2$, the bound (\ref{central}) is not a sufficient
condition it can be complemented by additional constraints based on 
partial traces of $H^{-1} I^N(M,\theta)$ which we now exhibit.

Consider a subset $i=1,\dots,p'$ ($p'<p$) of the parameters. Let $\rho_{,i'}$ be
the corresponding derivatives of $\rho(\theta)$. Let us define the
effective dimension $d'$ of the space in which these
parameters act at the point $\theta^0$ as follows. Let $\Pi$ be a
projector that commutes with $\rho(\theta^0)$ ($[\Pi,\rho(\theta^0)]=0$) 
and such that $\rho_{,i'}$, $i'=1,\dots,p'$ acts only
within the eigenspace of $\Pi$ 
(that is $\Pi \rho_{,i'}\Pi = \rho_{,i'}$). 
Then $d'$ is the smallest dimension of the eigenspace of such
a projector $\Pi$ ($d' = \Tr \Pi$).
To be more explicit, let 
us reexpress the definition of $d'$ in coordinates. First we
diagonalize $\rho(\theta^0) = \sum_k p_k |k\rangle\langle k|$. 
If some $p_k$ are equal this can be done in many ways.
The projector $\Pi$ projects onto some of the eigenvectors of $\rho$:
$\Pi = \sum_{k=1}^{d'} |k\rangle\langle k|$.
Next we write the operators $\rho_{,i'}$  in this basis: 
$\rho_{,i'} = \sum_{k,l=1}^{d'} (\rho_{,i'})_{kl} |k\rangle\langle l|$  
where the fact that the indices $k,l$ go from one to $d'$
expresses the fact that $ \rho_{,i'}$ acts only within the eigenspace
of $\Pi$. Finally we choose the smallest such $d'$.

We will  show that

\begin{equation}
\sum_{i',j'=1}^{p'} H^{ -1}_{i'j'} I^N_{i'j'} (M,\theta^0) \leq
N(d' -1)\ .
\label{extra1}
\end{equation}

Before proving this result let us illustrate it by an
example. Consider an unknown pure state in $d$ dimensions. In the
neighbourhood of a particular point we can parameterize the state by

\begin{equation}
\psi = \left|1\right\rangle + (\theta_2 + i \eta_2) 
\left|2 \right\rangle + ... + 
 (\theta_d + i \eta_d) \left|d \right\rangle
\end{equation}
 where the unknown parameters are
$\theta_i$ and $\eta_i$, $i=2,\dots,d$. There are thus $2d-2$ parameters.
At the point $\theta=\eta=0$, $H$ is diagonal in this
parameterization:
$H_{\theta_i \theta_j} = \delta_{ij}$, 
$H_{\eta_i \eta_j} = \delta_{ij}$,
$H_{\theta_i \eta_j} = 0$.
Hence (\ref{central}) takes the form 

\begin{equation}
\sum_i I^N_{\theta_i \theta_i}(M, \theta=\eta=0)
+ I^N_{\eta_i \eta_i}(M, \theta=\eta=0) \le N(d-1).
\label{pp}
\end{equation}
But using (\ref{extra1}) we also  find the constraints

\begin{equation}
I^N_{\theta_i \theta_i}(M, \theta=\eta=0)
+ I^N_{\eta_i \eta_i}(M, \theta=\eta=0) \leq N\quad , \quad i= 2,\dots,d
\end{equation}
which are stronger then (\ref{pp}) since they must hold
separately, but by summing them one obtains (\ref{pp}).

The proof of equation (\ref{extra1}) proceeds 
as in section \ref{Central}.
First we can restrict ourselves to POVM's whose elements are
proportional to one dimensional projectors. Secondly, we can 
restrict ourselves to the subspace $\Pi$ in 
evaluating (\ref{extra1}).
This follows from the inequality

\begin{eqnarray}
I(M)_{i'j'}
&=& \sum_\xi { \Tr (\rho_{,i'} M_\xi) \Tr (\rho_{,j'} M_\xi) \over
\Tr (\rho M_\xi)} \nonumber\\
&=&\sum_\xi { \Tr (\rho_{,i'}\Pi M_\xi\Pi) \Tr (\rho_{,j'}\Pi M_\xi\Pi) \over
\Tr (\rho\Pi M_\xi \Pi) + \Tr (\rho(1 -\Pi) M_\xi(1 - \Pi))} \nonumber\\
&\leq&
\sum_\xi { \Tr (\rho_{,i'}\Pi M_\xi\Pi) \Tr (\rho_{,j'}\Pi M_\xi\Pi) \over
\Tr (\rho\Pi M_\xi \Pi) }.
\label{cond}
\end{eqnarray}
Note that equality in (\ref{cond}) 
holds when the measurement consists of one
dimensional projectors and when the POVM decomposes into the sum of
two POVM's acting on the subspaces
spanned by $\Pi$ and $1- \Pi$ separately (i.e., the POVM elements
$M_\xi = \left| \psi_\xi\right\rangle\left\langle\psi_\xi\right|$ 
must commute 
with  $\Pi$ and $1- \Pi$).
Thirdly, we can increase the
number of parameters from $p'$ to $d'^2-1$.
We then introduce exactly as in (\ref{cshm}) 
a parameterization in which the $\rho_{,i}$ are
particularly simple, but in place of (\ref{rel}) we use

\begin{equation}
\sum_{1 \leq m' \leq d'}  c_{m'k'} c_{m'l'} = \delta_{k'l'} p_{k'} - 
{p_{k'} p_{l'} \over \Tr( {\Pi \rho})}.
\label{rel2}\end{equation}
After these preliminary steps the l.h.s.\ of
(\ref{extra1}) is calculated exactly as in subsections \ref{VB},
\ref{onemixedstate} and \ref{Nmixedstates}.

\section{Dropping the condition of unbiased estimators}\label{optimality}

\subsection{Quantum van Trees inequality}

In the previous section we proved a bound on the m.q.e.\ of unbiased
estimators $\hat \theta^N$ of $N$ copies of the quantum system 
$\rho(\theta)$ (with the additional condition that if $\rho$ is mixed
the measurement should be separable). 
In this section we shall prove Theorems IV and V
that under additional conditions it is possible to drop 
the hypothesis that the estimator is unbiased. 

The starting point for the results in this section is a Bayesian form
of the Cram\'er-Rao inequality, the van Trees inequality \cite{vanTrees}, 
and in particular the
multivariate form of the van Trees inequality proven in \cite{GillLevit}.
Adapted to the problem of estimating a quantum state, this inequality
takes the following form.
Let $\hat\theta^N$ be an arbitrary estimator of the parameter $\theta$
based on a measurement $M$ of the system $\rho^N(\theta)$.
Suppose it has m.q.e.\ matrix ${ V}^N(\theta)$,
and Fisher information matrix $I^N(M,\theta)$. 
Let $\lambda(\theta)$ be a smooth
density supported on a compact region (with smooth boundary) of the
parameter space, and suppose $\lambda$ vanishes on the boundary.
By $\mbox{E}_\lambda$ we denote expectation over 
a random parameter value
$\Theta$ with the probability density $\lambda(\theta)$. Let $C(\theta)$
and $D(\theta)$ be two $p\times p$ matrix valued functions of $\theta$,
the former being symmetric and positive definite. 
Then the multivariate van Trees inequality reads

\begin{equation}
\mbox{E}_\lambda\Tr C(\Theta) { V}^N(\Theta)\ge {{  (\mbox{E}_\lambda\Tr
                           D(\Theta))^2 }  \over
  { \mbox{E}_\lambda \Tr C(\Theta)^{-1} D(\Theta) I^N(M,\Theta)
      D(\Theta)^\top
   +  \tilde{\cal I}(\lambda)  }}
\label{vantrees}
\end{equation}
where $\ ^\top$ denotes the transpose of the matrix and

\begin{equation}
\tilde {\cal I}(\lambda )
= \int d \theta {1 \over \lambda (\theta )}
\sum_{ijkl} C_{ij}(\theta )^{-1}
\partial_{\theta_k}\{D_{ik}(\theta ) \lambda(\theta)\}
\partial_{\theta_l}\{D_{jl}(\theta ) \lambda(\theta)\}.
\label{itilde}
\end{equation}

As a first application of this inequality we shall prove Theorem V,
that is bound the
minimum value averaged over $\theta$ of a quadratic cost function.
Let $C(\theta)$ be the quadratic cost function. 
Consider the matrix $W_{opt}(\theta)$ that minimizes 
for each value of $\theta$ the cost $\Tr C(\theta) W(\theta)$ under
the condition that $\Tr H(\theta)^{-1} W(\theta)^{-1} \le d-1$.
One easily finds that

\begin{eqnarray}
W_{opt}&=& {\Tr \sqrt{ H^{-1/2} C H^{-1/2} } \over d-1 }
H^{-1/2} \sqrt{ H^{1/2} C^{-1} H^{1/2} }H^{-1/2}\\
&=&{\Tr \sqrt{ C^{1/2} H^{-1} C^{1/2} } \over d-1 }
C^{-1/2} \sqrt{ C^{1/2} H^{-1} C^{1/2} }C^{-1/2}
\label{Wopt}
\end{eqnarray}
and that

\begin{eqnarray}
\Tr C W_{opt} = {\left ( \Tr \sqrt{ H^{-1/2} C H^{-1/2} } \right)^2
\over d-1} = {\left ( \Tr \sqrt{ C^{1/2} H^{-1} C^{1/2} } \right)^2
\over d-1} \ .
\label{WoptC}
\end{eqnarray}

We choose in (\ref{vantrees}) $D(\theta) = C(\theta) 
W_{opt}(\theta) $. Thus $\Tr D(\theta) = \Tr  C(\theta) 
W_{opt}(\theta)$ is given by (\ref{WoptC}).
Note that

\begin{eqnarray}
 D(\theta)^\top C(\theta)^{-1} D(\theta) =
W_{opt}(\theta)  C(\theta)W_{opt}(\theta)
=  {\Tr  C(\theta)  W_{opt}(\theta) \over d-1} H(\theta)^{-1}.
\end{eqnarray}
Thus

\begin{eqnarray}
\Tr D(\theta)^\top C(\theta)^{-1} D(\theta)  I^N(M,\theta)
&=&  {\Tr  C(\theta)  W_{opt}(\theta) \over d-1} 
\Tr H(\theta)^{-1} I^N(M,\theta)\nonumber\\
&\leq& N \Tr  C(\theta)  W_{opt}(\theta) \ .
\end{eqnarray}

Inserting these expressions into (\ref{vantrees}) one obtains

\begin{eqnarray}
\mbox{E}_\lambda\Tr C(\Theta) { V}^N(\Theta)&\ge&
 {{  (\mbox{E}_\lambda\Tr C(\Theta)  W_{opt}(\Theta ))^2 }  \over
N \mbox{E}_\lambda\Tr C(\Theta)  W_{opt}(\Theta ) 
 +  \tilde{\cal I}(\lambda)  }\nonumber\\
&\ge& {  \mbox{E}_\lambda\Tr C(\Theta)  W_{opt}(\Theta )^2   \over
N} - {\alpha \over N^2}
\end{eqnarray}
where

\begin{equation}
\alpha = { \tilde{\cal I}(\lambda) \over
\mbox{E}_\lambda\Tr C(\Theta)   W_{opt}
(\Theta) }
\end{equation}
is independent of $N$.
This proves that upon averaging over $\theta$ it is impossible 
(for large $N$) to improve over the minimum cost (\ref{CVNopt}).

\subsection{Asymptotic version of the Cram\'er-Rao inequality}

We now prove Theorem IV, that is an asymptotic version  of our
main inequality (\ref{central2})  which is valid at every point $\theta$  and
does not make the assumption of unbiased estimators. 
We must however slightly restrict the class of
competing estimators since otherwise by the phenomenon of super-efficiency
we can beat a given estimator at any specific value of the parameter, though
we pay for this by bad behaviour closer and closer to the chosen value
as $N$ becomes larger. 

The restriction on the class of estimators is that $N$ times their
mean quadratic error matrix must converge uniformly in a neighbourhood
of the true value $\theta^0$ of $\theta$ to a limit $W(\theta)$, continuous at
$\theta^0$.
We assume that both $W(\theta^0)$ and $H(\theta^0)$ are nonsingular.
Furthermore we shall require some mild smoothness conditions on
$H(\theta)$ in a neighbourhood of $\theta^0$:
it must be continuous at $\theta^0$ with bounded partial derivatives
with respect to the parameter in a neighbourhood of $\theta^0$. Note
that imposing regularity conditions on $H$ is natural
since it corresponds to supposing that the $\theta_i$ smoothly
parametrize the allowed density matrices.

Suppose that as $N\to\infty$

\[ N {V}^N(\theta) \to W(\theta) \]
uniformly in $\theta$ in a neighbourhood of $\theta^0$, with
$W$ continuous at $\theta^0$; write $W^0=W(\theta^0)$.
Now in (\ref{vantrees}) let us make the following choices for the
matrix functions $C$ and $D$:

\[
C(\theta)={W^0}^{-1} H^{-1}(\theta) {W^0}^{-1},
\]

\[
D(\theta)={W^0}^{-1} H^{-1}(\theta).
\]
Then \eqref{vantrees} (multiplied throughout by $N$) and \eqref{itilde} become

\begin{eqnarray}
\mbox{E}_\lambda\Tr {W^0}^{-1} H^{-1}(\Theta) {W^0}^{-1}
    \,N{V}^N(\Theta)&\ge &
     {{  (\mbox{E}_\lambda\Tr
                           {W^0}^{-1} H^{-1}(\Theta))^2 }  \over
  { {1\over N}\mbox{E}_\lambda \Tr H^{-1} I^N(M,\Theta)
   +  {1\over N}\tilde{\cal I}(\lambda)  }} \nonumber\\
   &\ge&   {{  (\mbox{E}_\lambda\Tr
                           {W^0}^{-1} H^{-1}(\Theta))^2 }  \over
  { (d-1)
   +  {1\over N}\tilde{\cal I}(\lambda)  }}
\label{vantrees2}
\end{eqnarray}
and

\begin{equation}
\tilde {\cal I}(\lambda )
= \int d \theta {1 \over \lambda (\theta )}
\sum_{ijkl} H_{ij}(\theta )
\partial_{\theta_k}\{ H^{-1}_{ik}(\theta ) \lambda(\theta)\}
\partial_{\theta_l}\{{H}^{-1}_{jl}(\theta ) \lambda(\theta)\},
\label{itilde2}
\end{equation}
where we have used our central inequality (\ref{central}) to
pass to (\ref{vantrees2}). Now suppose that the quantity (\ref{itilde2})
is finite (we will give conditions for that in a moment). By
the assumed uniform convergence of $N{V}^N$ to $W$, upon letting
$N\to\infty$ (\ref{vantrees2}) becomes

\begin{equation}
\mbox{E}_\lambda\Tr {W^0}^{-1} H^{-1}(\Theta) {W^0}^{-1} W(\Theta)\ge
     {{  (\mbox{E}_\lambda\Tr
                           {W^0}^{-1} H^{-1}(\Theta))^2 }  \over
  { (d-1)  }}.
\label{vantrees3}
\end{equation}
Now suppose the density $\lambda$ in this equation
(the probability density of $\Theta$) is replaced by an
element $\lambda^m$
in a sequence of densities, concentrating on smaller and smaller neighbourhoods
of $\theta^0$ as $m\to\infty$. Assume that $H(\theta)$ is continuous
at $\theta^0$.
Recall our earlier assumption that $W(\theta)$ is also continuous at
$\theta^0$, with $W^0=W(\theta^0)$. Then taking the limit as $m\to\infty$
of (\ref{vantrees3}) yields

\[ \Tr W^{-1}(\theta^0)H^{-1}(\theta^0)\ge
  (\Tr W^{-1}(\theta^0)H^{-1}(\theta^0))^2/(d-1)
 \]
or the required limiting form of (\ref{central}),

\[
\Tr W^{-1}(\theta^0)H^{-1}(\theta^0)\le
  (d-1).
\]

It remains to discuss whether it was reasonable to
assume that $\tilde{\cal I}(\lambda^m)$ is finite (for each $m$ separately).
Note that this quantity only depends on the prior density $\lambda$ and
on $H(\theta)$, where $\lambda$ is one of a sequence of densities
supported by smaller and smaller neighbourhoods of $\theta^0$.
We already assumed that $H(\theta)$ was continuous at $\theta^0$.
It is certainly possible to specify prior densities $\lambda^m$
concentrating on the ball of radius $1/m$, say, satisfying the smoothness
assumptions in \cite{GillLevit} and with, for each $m$,
finite Fisher information matrix

\[
\int d \theta {1 \over \lambda^m (\theta )}
\partial_{\theta_k}\{\lambda^m(\theta)\}
\partial_{\theta_l}\{ \lambda^m(\theta)\}.
\]
Consideration of (\ref{itilde2}) then shows that it suffices further
just to assume that $\partial_{\theta_k}\{ H^{-1}_{ik}(\theta ) \}$
is, for each $i,k$, bounded in a neighbourhood of $\theta^0$.

In conclusion we have shown that under mild smoothness conditions on
$H(\theta)$, the limiting mean quadratic error matrix $W$
of a sufficiently regular but otherwise arbitrary sequence of estimators
must satisfy the asymptotic version of our
central inequality $\Tr H^{-1} W^{-1}\le d-1$.

\section{Attaining the Cram\'er-Rao bound in 2 dimensions}\label{2dim}

We shall now show that the bounds (\ref{central}), (\ref{central3}),
(\ref {central3b})
are sharp in the case of pure states of spin 1/2 systems and of
separable measurements in the case of mixed states of spin 1/2 systems. 
In particular, in the limit of a large number of
copies $N$ any target scaled m.q.e.\ matrix $W$ that 
satisfies $\Tr H^{-1}W^{-1}\leq 1$ can be attained 
(provided $W$ is non singular). 
We shall show this by explicitly constructing
a measurement strategy that attains the bound. 
In section \ref{optimality} we have already shown 
that if $\Tr H^{-1}W^{-1}> 1$, then it cannot be attained.

\subsection{Attaining the bound at a fixed point $\theta^0$}\label{2dtheta0}

The first step in the proof is to consider the case of one copy of the
unknown state ($N=1$) and fix a particular point $\theta^0$. 
Then we show that for any 
target information matrix $G(\theta^0)$ that
satisfies $\Tr H^{-1}(\theta^0) G (\theta^0) \leq 1$, we can build a 
measurement $M=M^{\theta^0}$, 
{\it in general depending on\/} $\theta^0$, 
such that $I(M^{\theta^0}, \theta^0) = G(\theta^0)$.
In the next sections  we shall show how to use this intermediate result
to build a measurement and estimation strategy 
whose asymptotic m.q.e.\ is equal to $W(\theta)=G(\theta)^{-1}$
for {\it all\/} $\theta$.

Let us first consider the case of pure states. At $\theta^0$, the state is
$|\psi^0\rangle$. We introduce a parameterization $\theta_1, \theta_2$ such
that in the vicinity of $|\psi^0\rangle$, the unknown state is

\begin{equation}
|\psi (\theta)\rangle = |\psi^0\rangle + (\theta_1 + i \theta_2)
|\psi^1\rangle
\ .
\end{equation}
Thus the original point $\theta^0$ corresponds 
to the new $\theta_1=\theta_2=0$.
In this parameterization, $H$ is proportional to the identity at
$\theta_1=\theta_2=0$: $H_{\theta_1 \theta_1}(0)= H_{\theta_2 \theta_2}(0)=1$, 
$ H_{\theta_1 \theta_2}(0)= 0$.

We now diagonalize the matrix $G$. Thus there exist new parameters
$\theta'_1= \cos \lambda \,\theta_1 + \sin \lambda \,\theta_2$, 
$\theta'_2= -\sin \lambda \,\theta_1 + \cos \lambda \,\theta_2$ such that
$G_{ \theta'_1 \theta'_1 }(0) = g_1\ge 0$, 
$G_{ \theta'_2 \theta'_2 }(0) = g_2\ge 
0$, $G_{ \theta'_1 \theta'_2 }(0) = 0$.

In terms of the parameters $\theta'_1, \theta'_2$, the unknown state 
is written

\begin{equation}\label{full pure}
|\psi^0\rangle = |\psi^0\rangle + (\theta'_1 + i \theta'_2) 
|{\psi^1}'\rangle
\end{equation}
where $|{\psi^1}'\rangle = e^{i \lambda}
|\psi^1\rangle$.

The POVM $M^{\theta^0}$ consists of
measuring the observable $|\psi^0\rangle\langle {\psi^1}'| +
|{\psi^1}'\rangle \langle \psi^0|$ with probability
$g_1$, of measuring  
the observable $i( |\psi^0\rangle\langle {\psi^1}'| -
|{\psi^1}'\rangle \langle \psi^0|)$ with probability
$g_{2}$, and of measuring nothing (or measuring the identity) with
probability $1 - g_1 - g_2$.
It is straightforward to verify that the Fisher 
information at $\theta^0$ in a measurement of the POVM $M^{\theta^0}$ is
equal to $G(\theta^0)$.

Let us now turn to the case of mixed states. We suppose that there
are three unknown parameters.
We use a parameterization in which
$\rho(\theta) = (1/2) (I + \theta \cdot \sigma)$,
with $\|\theta\|<1$.
Without loss of generality we can suppose that
$\theta^0 = (0,0,n)$, so that $\rho(\theta^0) = 
(1/2 + n/2) \left| 1\right\rangle\left\langle 1\right| + (1/2 -n/2) 
\left| 2\right\rangle\left\langle 2\right| = {1 \over 2}
(I + n \sigma_z)$.
The tangent space at $\rho$ is spanned by the Pauli matrices
$\rho_{,x} =  \sigma_x (={\rho_{,12+}\over 2}) $, $\rho_{,y} =  \sigma_y 
(={\rho_{,12-}\over 2})$,
$\rho_{,z} = \sigma_z  (=\rho_{,1}\sqrt{1 - n^2})$ where in
parenthesis we have given the relation to the basis used in section
\ref{onemixedstate}. 
In this coordinate system $H(\theta^0)$ 
is diagonal with eigenvalues $1$, $1$, $1/(1- n^2)$.

Take any symmetric positive matrix
$G$ satisfying $\Tr G H^{-1}(\theta^0) \le 1$. 
Define the matrix $F = H^{-{1 \over 2}} G H^{-{1 \over 2}}
= \sum_i \gamma_i f_i \otimes f_i$, where $\gamma_i$ and $f_i$ are the
eigenvalues and eigenvectors of $F$. The condition $\Tr G
H^{-1}(\theta^0) \le 1$ can then be rewritten $\sum_i \gamma_i \le
1$.
If we define $g_i = H^{{1 \over 2}} f_i$, then we can write $G=
\sum_i \gamma_i g_i \otimes g_i$. Denote $m_i = g_i/\|g_i\|$.

Consider the measurement of the spin along the direction $m_i$.
This is the POVM consisting of the two projectors
$P_{+m_i} = {1\over 2}(I + m_i. \sigma)$ and 
$P_{-m_i}  = {1\over 2}(I - m_i. \sigma)$.
The information matrix for this measurement is

\begin{equation}
I(P_{\pm m_i})_{kl} = \sum_\pm {\Tr (P_{\pm m_i} \sigma_k)\Tr (P_{\pm m_i}
\sigma_l)
\over \Tr (P_{\pm m} \rho)} = { (m_i)_k (m_i)_l \over (1 - n^2 (m_i)_z^2)}
\end{equation}
where $(m_i)_k$ is component $k$ of vector $m_i$.
Therefore this information matrix is proportional to $g_i\otimes
g_i$. One verifies 
that it obeys $\Tr H^{-1} I(P_{\pm
m_i}) =1$, as it must by our findings in section \ref{Central}
since the measurement is exhaustive, $N=1$, and $p=d^2-1$.
Therefore the coefficient of proportionality is $1$ and

\begin{equation}
I(P_{\pm m_i}) = g_i\otimes g_i \  .
\end{equation}

We now combine such POVM's to obtain the POVM whose elements are

\begin{equation}
\gamma_1 P_{+m_1}, \  \gamma_1 P_{-m_1}, \ 
\gamma_2 P_{+m_2}, \  \gamma_2 P_{-m_2}, \ 
\gamma_3 P_{+m_3}, \  \gamma_3 P_{-m_3}, \ 
(1 - \gamma_1 -\gamma_2-\gamma_3)\ .
\label{eighty}
\end{equation}
The information matrix for this measurement is just the sum
$\gamma_1 I(P_{\pm m_1}) + \gamma_2 I(P_{\pm m_2})
+ \gamma_3 I(P_{\pm m_3})= \sum_i \gamma_i g_i \otimes g_i = G$. 
Thus the POVM (\ref{eighty}) attains the target information
$G$ at the point $\theta^0$.

\subsection{Attaining the bound for every $\theta$ and arbitrary $N$
by separable measurements}\label{sub2}

We now prove Theorem VII that states that in the case of spin half particles
we can attain the bound (\ref{central3}) for every $\theta$.
Give yourself a continuous matrix $W(\theta)$, the target scaled m.q.e.\ matrix,
satisfying (\ref{central3}) for every $\theta$. Define $G(\theta)=W(\theta)^{-1}$,
the target scaled information matrix, which satisfies therefore (\ref{central}).
We will show that there exists a separable
measurement and an estimation strategy on
$N$ copies of the state $\rho(\theta)$
such that the m.q.e.\ matrix $V^N$ of the estimator satisfies

\begin{equation}V^N(\theta)_{ij}=
\mbox{E}_\theta((\hat \theta_i - \theta_i)(\hat \theta_j - \theta_j))
= {W_{ij}(\theta)  \over N} +o\left({1 \over N}\right)
\label{NN}
\end{equation}
for all $\theta$. In fact this holds uniformly in $\theta$ in a sufficiently
small neighbourhood of any given point.
This is proven by constructing explicitly a measurement and estimation
strategy that satisfies (\ref{NN}), following the lines of 
\cite{BarndorffNielsenGill}.

The measurement and estimation strategy we propose is the following:
first take a fraction $N_0=O(N^a)$ 
of the states, for some fixed $0<a<1$, and on one third of them measure 
$\sigma_x$, on one third $\sigma_y$ and on one third $\sigma_z$. 
One obtains from each measurement of $\sigma_x$
the outcome $\pm 1$ with probabilities ${1\over 2}(1\pm \theta_x)$,
and similarly for $\sigma_y$, $\sigma_z$.
Using this data we make a first estimate of $\theta$, 
call it $\tilde\theta$, for instance by equating the observed relative 
frequencies of $\pm 1$ in the three kinds of measurement to their theoretical 
values. If the state is pure this determines a first estimate of the
direction of polarization. If the state is mixed it is possible that
the initial estimate suggests that the Bloch vector lies outside the
unit sphere. This only occurs with exponentially small probability (in
$N_0$) and if this is the case the measurement is discarded. As
discussed below this only affects the mean quadratic error by $o(1/N)$.

On the remaining $N'=N-N_0$ states we carry out the measurement
$M=M^{\tilde\theta}$ such that 
$I(M^{\tilde\theta},\tilde\theta) = G(\tilde\theta)$
which we have just shown how to construct.
Note that $I(M^{\tilde\theta},\theta)= G(\theta)$ is only guaranteed
when $\theta$ is precisely equal to $\tilde\theta$. 
Write $I(M,\theta;\tilde\theta)$
for the Fisher information about $\theta$, based on the measurement
$M^{\tilde\theta}$ optimal at $\tilde\theta$, while the true value 
of the parameter is actually $\theta$. Given $\tilde\theta$,
each of the $N'$ second stage measurements represents one draw from the 
probability distribution 
$p(\xi|\theta;\tilde\theta) = \Tr M_\xi^{\tilde\theta} \rho(\theta)$. 
We use the 
classical m.l.e.\  based on this data only (with $\tilde\theta$ fixed at its 
observed
value) to estimate what is the value of $\theta$. Call this estimated value
$\hat \theta$. 

Let $\epsilon>0$ be fixed, arbitrarily small. Let $\theta^0$ denote
the true value of $\theta$. For given $\delta>0$ let $B(\theta^0,\delta)$
denote the ball of radius $\delta$ about $\theta^0$. Fix
a convenient matrix norm $\|\cdot\|$.
We have the exponential bound 

\begin{equation}\label{exp bound}
\Pr\{\tilde\theta \in B(\theta^0,\delta) \} \ge 1- C e^{-D N_0 \delta^2}
\end{equation}
for some positive numbers $C$ and $D$
(depending on $\delta$). The reason we take $N_0$ proportional to $N^a$ 
for some $0<a<1$ is that this ensures that $1- C e^{-DN_0}=o(1/N)$.

Modern results \cite{IbragimovHasminskii} on the m.l.e.\ $\hat\theta$ 
state that, under certain regularity conditions, the conditional 
m.q.e.\ matrix of $\hat\theta$  satisfies (at $\theta=\theta^0$, and 
conditional on $\tilde\theta$)

\begin{equation}
V^{N'}(\theta^0;\tilde\theta) = 
{ I(M,\theta^0;\tilde\theta)^{-1}\over N'}
+o\left({1\over N'}\right)\label{unif}
\end{equation} 
uniformly in $\theta^0$. We need however for the next step in our
argument that this same result 
is true uniformly in $\tilde\theta$ for given $\theta^0$. This could be 
verified by careful reworking of the proof in \cite{IbragimovHasminskii}.
Rather than doing that, we will explicitly calculate in subsection
\ref{mse} and \ref{cmeOM}
the conditional m.q.e.\  matrix of our estimator
and show that it satisfies (\ref{unif})
uniformly in $\tilde\theta$ in a small enough neighbourhood 
$B(\theta^0,\delta)$ of $\theta^0$. The `little $o$' in (\ref{unif}) 
refers to the chosen matrix norm.

We will also need that $I(M,\theta^0;\tilde\theta)^{-1}$ is
continuous in $\tilde\theta$ at $\tilde\theta=\theta^0$, at which
point it equals by our construction the target scaled m.q.e.\ 
$W(\theta^0)$. This is also
established in subsection \ref{mse}. Therefore, replacing if
necessary $\delta$ by a smaller value, we can guarantee
that  $I(M,\theta^0;\tilde\theta)^{-1}$ is within $\epsilon$
of $I(M,\theta^0;\theta^0)^{-1}=W(\theta^0)$ for all $\tilde\theta\in
B(\theta^0,\delta)$.

If $\tilde\theta$ is outside the domain $B(\theta^0,\delta)$, 
then the norm of $V^{N'}(\theta^0;\tilde\theta)$
is bounded by a constant $A$ since $\theta$ belongs to a compact domain.

Putting everything together we find that

\begin{eqnarray*}
\| N' V^N(\theta^0) - W(\theta^0) \|
&=&\left\|\int\left( 
N' V^{N'}(\theta^0;\tilde\theta)- W(\theta^0) \right)
dP(\tilde\theta)\right\| \\
&\leq & \int_{B(\theta^0,\delta)} 
\|N' V^{N'}(\theta^0;\tilde\theta)- 
                 W(\theta^0)\|dP(\tilde\theta) + A N' C' e^{-DN_0}\\
&=&\int_{B(\theta^0,\delta)}
    \| I(M,\theta^0;\tilde\theta)^{-1} +o(1) -
W(\theta^0)\|dP(\tilde\theta) 
    + o(1)\\
&\leq&\epsilon+o(1)+o(1).
\end{eqnarray*}
It follows since $N'/N\to 1$ as $N\to\infty$ that 
$\mathop{\lim\sup}\limits_{N\to\infty}
    \| N V^N(\theta^0) - W(\theta^0) \|  \le  \epsilon$. 
Since $\epsilon$ was arbitrary, we obtain (\ref{NN}).

\subsection{Analysis of the conditional mean quadratic error}\label{mse}

We first consider the case of impure states, with the parameterization

\begin{equation}\label{full model}
\rho={1\over 2}(I+\theta.\sigma),\quad\hbox{with }\sum(\theta_i)^2 < 1.
\label{impureP}
\end{equation}
where we have imposed that the state is never pure.
This case turns out to allow the most explicit and straightforward
analysis because the relation between the frequency of the outcomes
and the parameters $\theta$ is linear. 
For other cases the analysis is more delicate and is discussed in the
next subsection. In general,
smoothness assumptions will have to be made on the parameterization
$\rho=\rho(\theta)$. 

We suppose that $W(\theta)$ is non-singular and continuous in $\theta$. 
Consequently the $\gamma_i$ (defined in section \ref{2dtheta0})
depend continuously on $\theta$ and
are all strictly positive at the true value $\theta^0$ of $\theta$.

Given the initial estimate, the second stage measurement
can be implemented as follows: for each of the $N'=N-N_0$ observations, 
independently of one another, with
probability $\gamma_i$ measure the projectors $P_{\pm m_i}$, in other
words, measure the spin observable $m_i.\sigma$. 
With probability $1-\sum \gamma_i$ do nothing. 

We emphasize that the $\gamma_i$ and $m_i$ all depend on the initial
estimate $\tilde\theta$ through $W(\tilde\theta)$ and $H(\tilde\theta)$.
In the following, all probability calculations are 
conditional on a given value 
of $\tilde\theta$.

For simplicity we will modify the procedure in the following two ways: 
firstly, rather than
taking a random number of each of the three types of measurement, we
will take the fixed (expected) numbers $\lfloor \gamma_i N'\rfloor$ 
(and neglect the difference between $\lfloor \gamma_i N'\rfloor$ and 
$\gamma_i N'$). Secondly, we will ignore the
constraint $\sum(\theta_i)^2\le 1$. These two modifications make
the maximum likelihood estimator easier to analyze, but do not change its 
asymptotic m.q.e.  Later we will 
sketch how to extend the calculations
to the original {\it constrained\/} maximum likelihood estimator based on 
{\it random\/} numbers of measurements of each observable.

Now measuring $m_i.\sigma$ produces the values $\pm 1$ with probabilities
$p_{\pm i}={1\over 2}(1 \pm \theta.m_i)$. Since our data consists of
three binomially distributed counts and we have three parameters $\theta_1,
\theta_2,\theta_3$ the maximum likelihood estimator can be described,
using the invariance of maximum likelihood estimators under $1$--$1$ 
reparameterization, as follows:
set the theoretical values $p_{\pm i}$ equal to their empirical counterparts
(relative frequencies of $\pm 1$ in the $\gamma_i N'$ observations of the
$i$'th spin) and solve the resulting three equations in three unknowns.

To be explicit, define 
$\eta_i=2p_{+i}-1=\theta.m_i$ and let $\hat\eta_i$ be its empirical
counterpart. Recall that $m_i=g_i/\|g_i\|$, $g_i=H^{1/2} f_i$, where
the $f_i$ are the orthonormal eigenvectors of $H^{-1/2} G H^{-1/2}$,
and where $H$ and $G$ are $H(\tilde\theta)$, $G(\tilde\theta)$,
and $\tilde\theta$ is the preliminary estimate of $\theta$.
Then we can rewrite 

\[
\eta_i=\theta.m_i=
\theta.g_i/\|g_i\|=\theta.H^{1/2}f_i/\|H^{1/2}f_i\|
=(H^{1/2}\theta).f_i/\|H^{1/2}f_i\|
\]
from which we obtain

\[
(H^{1/2}\theta).f_i=\|H^{1/2}f_i\|\eta_i
\]
and hence

\[
\theta= H^{-1/2} \sum_i  \|H^{1/2}f_i\|\eta_i  f_i.
\]
The same relation holds between $\hat\theta$ and $\hat\eta$
and yields the sought for expression for $\hat\theta$ in
terms of the empirical relative frequencies.

Observing that the $\hat\eta_i$ are independent with variance 
$4p_{+m_i}p_{-m_i}/(\gamma_i N')=(1-(\theta.m_i)^2)/(\gamma_i N')$,
the m.q.e.\ matrix of $\hat\theta$, conditional on
the preliminary estimate $\tilde\theta$, is

\begin{equation}
V^{N'}(\theta^0;\tilde\theta)=
{1\over N'} \sum_i {1\over\gamma_i} \left( 
   1 - {{ (\theta^0.H^{1/2}f_i)^2 } \over { \|H^{1/2} f_i\|^2 }} \right)
      \|H^{1/2} f_i\|^2  H^{-1/2} (f_i\otimes f_i) H^{-1/2}.
\label{answer}
\end{equation}
There is no $o(1/N')$ term here so we do not have to check uniform
convergence: the limiting value is attained exactly. Actually
we cheated by replacing $\lfloor \gamma_i N'\rfloor$ by $\gamma_i N'$.
This does introduce a $o(1/N')$ error into (\ref{answer}) uniformly in a
neighbourhood of $\theta^0$ in which the $\gamma_i$, 
which depend on $\tilde\theta$, are bounded away from zero, and $H$
and its inverse are bounded.

One may verify that (\ref{answer}) reduces to $W(\theta^0)/N'$
at $\tilde\theta=\theta^0$ 
(indeed at $\theta^0=\tilde \theta$, $(\tilde \theta.H^{1/2}f_i)^2
= {n^2 f_{iz}^2 \over 1-n^2}$ and $\|H^{1/2} f_i\|^2= 
{1 - n^2 + n^2 f_{iz}^2\over 1-n^2}$).
But this computation is really
superfluous since at this point, we are computing the m.q.e.\ 
of the maximum likelihood estimator based on a measurement with,
by our construction, Fisher information equal to the inverse of $W(\theta^0)$.
(The modifications to our procedure do not alter the Fisher information).
The two quantities must be equal by the classical large sample results
for the maximum likelihood estimator.

We finally need to show the
continuity in $\tilde \theta$ at $\tilde\theta=\theta^0$ of $N'$ 
times the quantity in (\ref{answer}). 
This is evident if the $\gamma_i$ are all
different at $\theta^0$. Both the eigenvalues and the eigenvectors
of $H^{-{1\over 2}}GH^{-{1\over 2}}$ are then continuous functions of 
$\tilde\theta$ at $\theta^0$.
There is a potential difficulty however if some $\gamma_i$ are
equal to one another at $\tilde\theta=\theta^0$.
In this case, the eigenvectors $f_i$
are not continuous functions of $\tilde\theta$ at this point, and not
even uniquely defined there. 
We argue as follows that this does not destroy continuity of the mean
quadratic error.
Consider a sequence of points $\tilde\theta^n$ approaching $\theta^0$.
This generates a sequence of eigenvectors $f_i^n$ and 
eigenvalues $\gamma_i^n$.
The eigenvalues converge to the $\gamma_i$ but the eigenvectors need not
converge at all. However by compactness of the set of unit vectors in $R^3$, 
there is a subsequence along which the eigenvectors
$f_i^n$ converge; and they must converge to a possible choice of
eigenvectors at $\theta^0$. 
Thus along this subsequence
the mean quadratic error (\ref{answer}) does converge to a limit given 
by the same formula evaluated at the limiting $f_i$ etc. But this
limit is equal by construction to the inverse 
of the target information matrix $G(\theta)$. 
A standard argument now shows that the limiting mean quadratic error 
is continuous at $\tilde\theta=\theta^0$.

The m.q.e.\ of $\hat\theta$ given $\tilde\theta$ (times $N'$) 
therefore converges uniformly in a sufficiently small neighbourhood of
$\theta^0$ to a limit continuous at that point and equal to $W(\theta^0)$
there.

In our derivation of (\ref{NN}) we required the parameter 
and its estimator to be
bounded. By dropping the constraint on the length of $\theta$ we have
inadvertently lost this property. Suppose we replace our 
modified estimator $\hat\theta$ by the actual maximum likelihood estimator 
respecting the constraint. The two only differ when the unconstrained 
estimator lies outside the unit sphere; but this event only occurs
with an exponentially small probability, uniformly in $\tilde\theta$,
provided the $\gamma_i$ are uniformly bounded away from
$0$ in the given neighbourhood of $\theta^0$. From this it
can be shown that the mean quadratic error is altered by an amount $o(1/N')$
uniformly in $\tilde\theta$.

If we had worked with random numbers of measurements of each
spin variable, when computing the mean quadratic error
we would first have copied the computation above conditional on the numbers
of measurements, say $X_i$, of each spin $m_i$. These numbers are
binomially distributed with parameter $N'$ and $\gamma_i$.
The conditional mean quadratic error would be the same as the expression above
but with $1/(\gamma_i N')$ replaced by $1/X_i$ (and special provision
taken for the possible outcome $X_i=0$). So to complete the argument
we must show that $\mbox{E}(1/X_i)=1/(\gamma_i N')+o(1/N')$
uniformly in $\tilde\theta$. 
This can also be shown to be true, using the fact that $X_i/N'$ only differs 
from its mean by more than a fixed amount with exponentially small probability 
as $N'\to\infty$ and we restrict attention to $\tilde\theta$ in a neighbourhood 
of $\theta^0$ where the $\gamma_i$ are bounded away from zero.

Inspection of our argument shows that the convergence of the mean quadratic
error is uniform in $\theta^0$ as long as we keep away from the boundary
of the parameter space.

By the convergence of the normalized binomial distribution
to the normal distribution, the representation of the estimator we gave 
above also shows that
it is asymptotically normally distributed with asymptotic covariance
matrix equal to the target covariance matrix $W$. Moreover,
if $X$ has the binomial$(n,p)$ distribution, then $n^{1\over 2}(X/n-p)$
converges in distribution to the normal with mean zero and variance
$p(1-p)$, uniformly in $p$. Thus the
convergence in distribution is also uniform in $\theta^0$ as long as we 
keep away from the boundary of the parameter space.

\subsection{Conditional mean quadratic error for other models}\label{cmeOM}

The preceding subsection gave a complete analysis of the mean quadratic
error, given the preliminary estimate $\tilde \theta$ for the 3
unknown parameters $\theta_j$ of the parameterization (\ref{impureP}). 
We shall first analyze the mean quadratic error when the unknown
parameters are functions $\phi_i(\theta_j)$ of the parameters
$\theta_j$. We shall then consider the important case when the state is
pure and depends on two unknown parameters, and finally the case when
the state is pure or mixed and depends on one unknown parameter, or is 
mixed and depends on two unknown parameters.

Our first result is that if the change of parameters 
$\phi_i(\theta_j)$ is locally $C^1$, then the m.q.e.\ matrix of the
$\phi_i$ is obtained from the m.q.e.\ of the $\theta_j$ by the Jacobian
$\partial \phi_i / \partial \theta_j$ except eventually at isolated
points. This follows from the fact that
under a smooth (locally $C^1$) parameterization, 
the delta method (first order Taylor
expansion) allows us to conclude uniform convergence 
of the probability distribution of $\sqrt{N} ( \widehat \phi^N -\phi)$
to a normal limit with the target mean quadratic error. 
If the $\phi_i$ and their  derivatives  $\partial \phi_i / \partial \theta_j$ 
are bounded then this proves our  claim.
If there are points where the $\phi_i$ or their derivatives  
 $\partial \phi_i / \partial \theta_j$ are infinite, then 
convergence in distribution
does not necessarily imply convergence of moments. However a truncation
device allows one to modify the estimate $\widehat\phi$, replacing
it by $0$ if any component is larger than $c N^a$ for given $c$ and $a$
(use the method of \cite{IbragimovHasminskii}, Lemma II.8.2 together with
the exponential inequality \eqref{exp bound} for the multinomial 
distribution).
With this minor modification 
one can show (uniform in $\phi$ in a neighbourhood of $\phi^0$) 
convergence of the moments of the corresponding 
$\sqrt N (\widehat\phi-\phi)$ to the moments of its limiting distribution,
hence achievement of the bound in the sense of Theorem IV.
In particular if the parameter $\phi$ is bounded then the truncation is
superfluous.

Now turn to the pure state analog of model \eqref{full model}.
Obtain a preliminary estimate of the location of $\rho$ on the surface 
of the Poincar\'e sphere using the same method as in the mixed case, 
but always projecting onto the surface of the sphere. Next,
after rotation to transform the preliminary estimate into `spin up',
reparameterize to $\rho=\frac 12(1+\phi\cdot\sigma)$ where the parameters
to be estimated are  $(\phi_1,\phi_2)=(\theta'_1,\theta'_2)$ 
of the parameterization \eqref{full pure} while
$\phi_3=\sqrt{}(1- \phi_1^2-\phi_2^2)$. The preliminary estimate is
at $\phi_1=\phi_2=0$. The optimal measurement at this point according
to Section \ref{2dtheta0}
 consists of measurements of the spins $\sigma_1$ and $\sigma_2$ on 
specified proportions of the remaining copies. The
resulting estimator of the parameter $(\phi_1,\phi_2)$ is a linear
function of binomial counts and hence its mean quadratic error can be
studied exactly as in section \ref{mse}. Then we must transfer
back to the originally specified parameterization, for instance polar
coordinates.  This is done as in the preceeding paragraph. 
If the transformation
is locally $C^1$ then uniform convergence in distribution to the normal law
also transfers back; convergence of mean quadratic error too if the
original parameter space is bounded. Otherwise a truncation might be
necessary. In any case, we can exhibit a procedure optimal in the sense 
of Theorem  IV.

It remains to consider one- and two-dimensional sub-models of the
full mixed model, and one-dimensional sub-models of the full pure model.
We suppose that the model specifies a smooth curve or surface in the
interior of the Poincar\'e sphere, or a smooth curve on its surface;
smoothly parameterized by a one- or two-dimensional parameter as
appropriate. The first stage of the procedure is just as before, 
finishing in projection of an estimated density matrix into the model.
Then we reparameterize locally, augmenting the dimension of the
parameter to convert the model into a full mixed or pure model
respectively. The target information for the extra parameters is zero.
Compute as before the optimal measurement at this point. Because of
the zero values in the target information matrix, there will be zero
eigenvalues $\gamma_i$ in the computation of section \ref{2dtheta0}. Thus
the optimal measurement will involve specified fractions of measurement
of spin in the same number of directions as the dimension of the model.
Compute the maximum likelihood estimator of the original parameters based
on this data. If the parameterization is smooth enough the estimator will
yet again achieve the bound of Theorem IV.

\section{Conclusions and Open Questions}\label{conclude}

In this paper we have solved some of the theoretical problems that arise when
trying to estimate the state of a quantum system of which one
possesses a large number of copies.
This constitutes a preliminary step towards solving
the question with which Helstrom concluded his book \cite{Helstrom}:
``(\dots)\ mathematical statisticians
are often concerned with asymptotic properties of decision strategies and
estimators. (\dots)\ When the parameters of a quantum
density operator are estimated on the basis of many observations, how does the
accuracy of the estimates depend on the number of observations as that number
grows very large? Under what conditions have the estimates asymptotic
normal distributions? Problems such as these, and still others that
doubtless will occur to physicists and mathematicians, remain to be
solved within the framework of the quantum-mechanical theory.''

In the case of pure states of spin 1/2 particles the problem has been
solved. The key result is that in the limit of large $N$, the variance of the
estimate is bounded by (\ref{central2}), and the bound can be attained
by separate von Neumann measurements on each particle.

In the case of mixed states of spin 1/2 particles the state estimation
problem for large $N$ has been solved if one restricts oneself to
separable measurements. However if one considers non separable
measurements, then one can improve the quality of the estimate, which
shows that the Fisher information, which in classical statistics is
additive, is no longer so for quantum state estimation.

For the case of mixed states of spin 1/2
particles, or for
higher spins we do not know what the ``outer'' boundary of
the set of (rescaled) achievable Fisher information matrices based on 
arbitrary (non separable) 
measurements of $N$ systems looks like. We have some indications about
the shape of this set (see section \ref{Comp}) and we know that it is 
convex and compact.

\acknowledgements

S.M. thanks Utrecht University, Netherlands, where part of this work
was carried out. He is a {\em chercheur qualifi\'e } of the Belgium
National Research Fund (FNRS). R.D.G. thanks the generous hospitality 
of the Department of
Mathematics and Statistics, University of Western Australia.

\end{document}